\documentclass[twocolumn]{aastex631}

\let\pwiflocal=\iffalse \let\pwifjournal=\iffalse
\usepackage[utf8]{inputenc}
\usepackage{textcomp}
\usepackage{amsmath,amssymb,gensymb,times,graphicx,morefloats,color}
\usepackage{afterpage, bm}
\usepackage{natbib,hyperref}
\usepackage[outdir=./]{epstopdf}
\usepackage{mathrsfs}
\usepackage{mathtools}
\usepackage{url}
\usepackage{rotating}
\newcommand{\rstar}{$R_\star$}
\newcommand{\mstar}{$M_\star$}

\newcommand{\lsun}{$L_\odot$}

\newcommand{\mdot}{$\dot{M}$}

\newcommand{\hi}{H {\scshape i}}

\newcommand{\spitzer}{{\it Spitzer}}

\newcommand{\jwst}{{\it JWST}}

\newcommand{\lacc}{$L_{\rm acc}$}
\newcommand{\up}{{\it u}$^\prime$}
\newcommand{\rp}{{\it r}$^\prime$}

\graphicspath{{./}{}}

\received{Feb 07, 2025}
\revised{Mar 31, 2025}
\accepted{Apr 09, 2025}

\shortauthors{Tofflemire et al.}
\shorttitle{Accretion Diagnostics for \jwst--MIRI}

\begin{document}

\title{Coordinated Space and Ground-Based Monitoring of Accretion Bursts in a Protoplanetary Disk: \\ Establishing Mid-Infrared Hydrogen Lines as Accretion Diagnostics for \jwst--MIRI}
\thanks{Based on observations collected at the European Organization for \\ Astronomical Research in the Southern Hemisphere under ESO \\ program 114.2799.001.}

\correspondingauthor{Benjamin M.\ Tofflemire}
\email{btofflemire@seti.org}

\author[0000-0003-2053-0749]{Benjamin M.\ Tofflemire}
\affiliation{SETI Institute, 339 Bernardo Ave., Suite 200, Mountain View, CA 94043, USA}
\affiliation{NASA Ames Research Center, Moffett Field, CA 94035 USA}
\affiliation{Department of Astronomy, The University of Texas at Austin, Austin, TX 78712, USA}

\author[0000-0003-3562-262X]{Carlo F.\ Manara}
\affiliation{European Southern Observatory, Karl-Schwarzschild-Strasse 2, 85748 Garching bei München, Germany}

\author[0000-0003-4335-0900]{Andrea Banzatti}
\affiliation{Department of Physics, Texas State University, 749 North Comanche Street, San Marcos, TX 78666, USA}

\author[0000-0001-7552-1562]{Klaus M.\ Pontoppidan}
\affiliation{Jet Propulsion Laboratory, California Institute of Technology, 4800 Oak Grove Drive, Pasadena, CA 91109, USA}

\author[0000-0002-5758-150X]{Joan Najita}
\affiliation{NSF's NOIRLab, 950 North Cherry Avenue, Tucson, AZ 85719, USA}

\author[0000-0002-9190-0113]{Brunella Nisini}
\affiliation{INAF–Osservatorio Astronomico di Roma, Via di Frascati 33, 00078, Monte Porzio Catone, Italy}

\author[0000-0002-3741-9353]{Emma T.\ Whelan}
\affiliation{Department of Physics, Maynooth University, Maynooth, Co. Kildare, Ireland}

\author[0000-0002-3913-3746]{Justyn Campbell-White} 
\affiliation{European Southern Observatory, Karl-Schwarzschild-Strasse 2, 85748 Garching bei München, Germany}

\author[0009-0002-4535-1704]{Hala Alqubelat} 
\affiliation{European Southern Observatory, Karl-Schwarzschild-Strasse 2, 85748 Garching bei München, Germany}

\author[0000-0001-9811-568X]{Adam L.\ Kraus}
\affiliation{Department of Astronomy, The University of Texas at Austin, Austin, TX 78712, USA}

\author[0000-0003-1817-6576]{Christian Rab}
\affiliation{University Observatory, Faculty of Physics, Ludwig-Maximilians-Universität München, Scheinerstr. 1, D-81679 Munich, Germany}
\affiliation{Max-Planck-Institut für extraterrestrische Physik, Giessenbachstrasse 1, D-85748 Garching, Germany}

\author[0000-0001-8790-9011]{Adrien Houge} 
\affiliation{Center for Star and Planet Formation, GLOBE Institute, University of Copenhagen, Øster Voldgade 5-7, DK-1350 Copenhagen, Denmark}

\author[0000-0002-3291-6887]{Sebastiaan Krijt}
\affiliation{Department of Physics and Astronomy, University of Exeter, Exeter, EX4 4QL, UK}

\author[0000-0002-5943-1222]{James Muzerolle}
\affiliation{Space Telescope Science Institute, 3700 San Martin Dr, Baltimore, MD 21218, USA}

\author[0000-0002-5261-6216]{Eleonora Fiorellino}
\affiliation{Instituto de Astrofísica de Canarias, IAC, Vía Láctea s/n, 38205 La Laguna (S.C.Tenerife), Spain}
\affiliation{Departamento de Astrofísica, Universidad de La Laguna, 38206 La Laguna (S.C.Tenerife), Spain}

\author[0000-0002-7695-7605]{Myriam Benisty}
\affiliation{Max-Planck Institute for Astronomy (MPIA), Königstuhl 17, 69117 Heidelberg, Germany}

\author[0000-0002-9470-2358]{Lukasz Tychoniec}
\affiliation{Leiden Observatory, Leiden University, 2300 RA Leiden, The Netherlands}

\author[0000-0003-3682-6632]{Colette Salyk}
\affiliation{Department of Physics and Astronomy, Vassar College, 124 Raymond Avenue, Poughkeepsie, NY 12604, USA}

\author{Guillaume Bourdarot}
\affiliation{Max-Planck-Institut für extraterrestrische Physik, Giessenbachstrasse 1, D-85748 Garching, Germany}

\author[0009-0008-2382-0614]{Jacob Hyden} 
\affiliation{Department of Physics, Texas State University, 749 North Comanche Street, San Marcos, TX 78666, USA}

\begin{abstract}
In this paper, we establish and calibrate mid-infrared hydrogen recombination lines observed with \jwst\ as accretion tracers for pre-main-sequence stars that accrete from circumstellar disks. This work is part of a coordinated, multi-observatory effort that monitored the well-known binary system DQ Tau over three orbital periods, capturing its periodic accretion bursts.
In this first paper, we present 9 epochs of MIRI-MRS spectra with near-simultaneous LCO photometry and VLT--X-Shooter spectroscopy. This program caught exceptional accretion variability, spanning almost two orders of magnitude between the peak of the first periastron accretion burst and the following quiescent phases. The MIRI spectra show \hi\ line luminosities that vary in step with the accretion-luminosity time series measured with LCO and X-Shooter. The tight correlation with accretion and the large line widths, which MIRI resolves for the first time, support an accretion-flow origin for mid-infrared \hi\ transitions. Combining these three exceptional datasets, we derive accurate relations between mid-infrared line and accretion luminosities for three \hi\ transitions (10-7, 7-6, 8-7), and improve upon a previous relation based on Spitzer spectra. These new relations equip the community with a direct measurement of the accretion luminosity from MIRI-MRS spectra. A MIRI-derived accretion luminosity is fundamental for time-domain chemistry studies, as well as for studies of accretion in embedded/distant sources that are currently inaccessible in the optical. With these new relations, we provide accretion luminosities for an archival sample of 38 MRS spectra of protoplanetary disks published to date.
\end{abstract}

\section{Introduction} 

Protoplanetary disks hold the raw material for planet formation. The evolution of a disk’s mass and chemistry sets the timescale and conditions for planet formation \citep[e.g.,][]{Miotello2023}, as well as the composition of planets and their atmospheres \citep[e.g.,][]{ObergBergin2021,Drazkowska2023}. A primary ingredient for both mass and chemical evolution is the accretion rate onto the central star \citep[e.g.,][]{Hartmannetal2016}. As accretion consumes disk material and drives outflows, exhausting the disk, accretion shocks at the stellar surface bathe the disk in ionizing radiation, driving rich photo-chemistry \citep[e.g.,][]{Oberg2023,vanDishoeck2023}. A measure of the accretion rate is integral to the detailed characterization of a disk, and in the inference of its future or past \citep[e.g.,][]{ManaraetalPPVII}. 

The mass accretion rate is measured by modeling the radiation released by the process. In the magnetospheric accretion paradigm, stellar magnetic fields truncate the inner disk, confining material to flow along fields lines until it impacts the stellar surface \citep{Hartmannetal1994,Shuetal1994}. A standing shock emits from optically thick and thin regions at $\sim 10^4$ K, producing continuum and line emission \citep{Hartmannetal2016}. The hot thermal and Balmer continua make the signatures of accretion strongest in the near-ultraviolet (NUV), and readily apparent as an excess above the typically faint photospheres of low mass, T-Tauri stars.

Flux calibrated, medium-resolution spectra in the NUV and optical offer the best resource for measuring the accretion luminosity. In these spectra, the continua of accreting sources can be well-fit with the combination of a non-accreting stellar template, an accretion model (for one or more accretion columns), and an extinction law \citep[e.g.,][]{Valentietal1993,Calvet&Gullbring1998,HerczegHillenbrand2008,Manaraetal2013fitter,Inglebyetal2013,Espaillatetal2022}. When joined with stellar parameters (\mstar, \rstar), a mass accretion rate can be computed. From these measurements, a host of accretion diagnostics have been established by calibrating relations with broadband NUV photometry \citep[e.g.,][]{Gullbringetal1998,Robbertoetal2004}, emission line luminosities, \citep[e.g.,][]{Mohantyetal2005,HerczegHillenbrand2008,Fangetal2009,Alcalaetal2014,Alcalaetal2017}, or the \hi\ Balmer $\alpha$ line width \citep[e.g.,][]{Nattaetal2004}. Often, these relations are made from observations that are not made simultaneously, introducing scatter from the intrinsic variability of the accretion process. With few exceptions, these diagnostics reside in the NUV, optical, and near-infrared (NIR) wavelengths. 

As \jwst\ reopens access to mid-infrared (MIR) spectroscopy, the need for MIR accretion diagnostics is twofold. First, \jwst’s sensitivity allows for the study of faint and embedded sources that are not amenable to observation at shorter wavelengths \citep[e.g.,][]{Fiorellino2023,LeGouellec2024}\footnote{We note that for some highly embedded sources, scattered light from the envelope may be a source of contamination to contend with even for MIR \hi\ lines.}. For many sources, \jwst\ may be the only facility capable of conducting observations. The characterization of younger (more embedded) disks and larger, more diverse disk populations (more distant) requires MIR accretion tracers. 

Second, the rich forest of molecular lines in the MIR enables detailed chemical studies of the inner regions of protoplanetary disks \citep[e.g.,][]{banzatti23,grant23,Tabone2023,Colmenares2024,Kanwar2024}, where accurate photochemical modeling may greatly benefit (or require) a simultaneous measurement of the accretion luminosity. Accretion is a notoriously stochastic process giving rise to variability on timescales from hours to decades \citep{FischeretalPPVII}. The impact variable accretion has on inner disk chemistry has only been detected under extreme conditions (i.e., episodic accretion bursts, e.g. in EX Lup; \citealt{Aspinetal2010,Banzattietal2012,banzatti15,Kospaletal2023,Woitke2024}), but it still eludes us in the more typical accretion variability common to the T~Tauri phase \citep{banzatti14}. As chemical studies of protoplanetary disks enter the time domain, simultaneous accretion measurements will be increasingly fundamental to reveal how accretion variability affects the chemistry of planet-forming regions. 

\begin{figure*}
\centering
\includegraphics[width=1\textwidth]{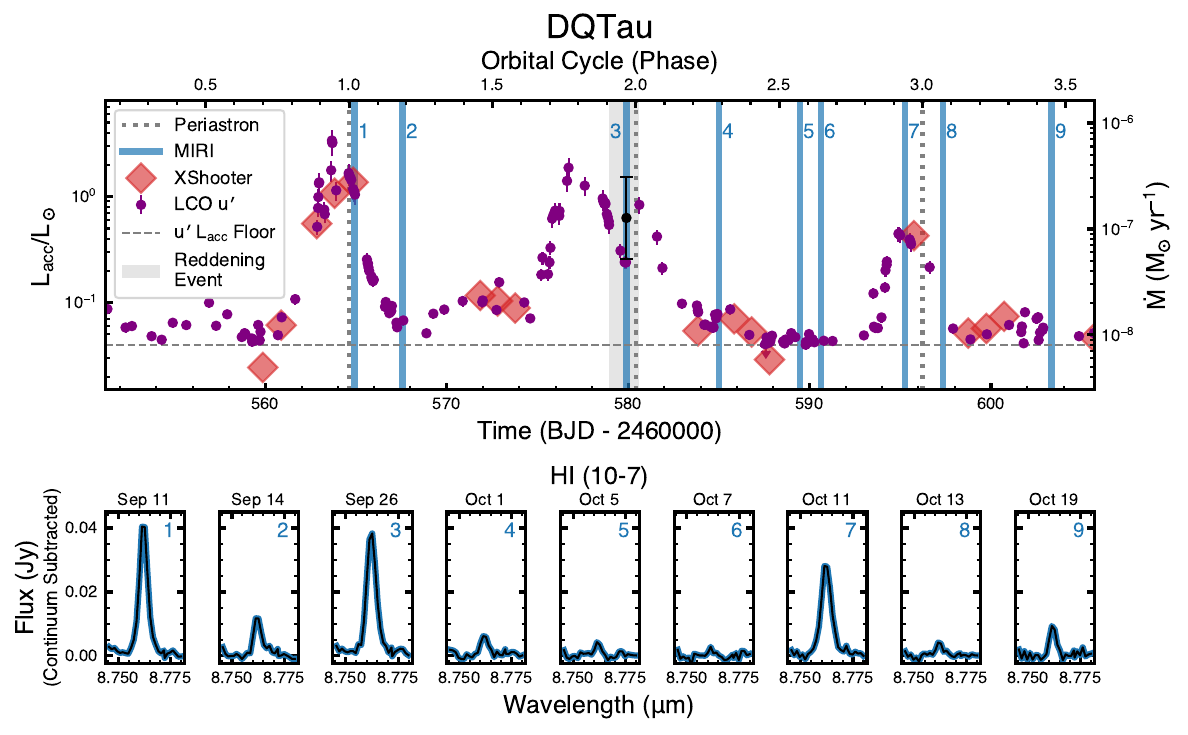}
\caption{DQ Tau accretion variability. {\bf Top:} The accretion luminosity (left axis) and mass accretion rate (right axis) from X-Shooter spectra and LCO \up-band photometry. Time is presented in barycentric Julian days on the bottom axis and the binary orbital cycle on the top axis. Vertical dotted lines mark binary periastron passages. Blue vertical lines mark the times of MIRI-MRS observations with an associated epoch number. The dashed horizontal line represents the accretion luminosity measurement floor for the \up\ photometry (see Section \ref{lco_lacc}; Figure \ref{fig: lacc_relation}). The gray vertical region surrounding MIRI Epoch 3 highlights a reddening event. The black point is an estimate of the accretion luminosity during this event (see Section \ref{reddening}).
{\bf Bottom:} Variability in the \hi\ (10-7) emission line. Continuum subtracted MIRI-MRS spectra centered on the line are presented for each observation, with the epoch number listed on the top right corner of each panel. The \hi\ (10-7) temporal behavior tightly tracks the densely sampled accretion luminosity measurements.}
\label{fig: lightcurve_epochs}
\end{figure*}

The most obvious choice for MIR accretion diagnostics are Hydrogen lines from high-order series (Humphreys series with n=6, and higher) that trace the cascade of recombination transitions in the accretion shock's optically thin emitting region. Previous work in the MIR introduced the \hi\ Pfund $\beta$ line at 4.65 $\mu$m, which has been commonly observed in ground-based $M$-band surveys. However, this line requires a complex correction for stellar photospheric absorption and disk CO emission that is only possible at high resolving powers \citep{salyk13}, making it difficult to use for accretion measurement with the current space-based instrumentation. Further in the MIR, \citet{rigliaco2015} used \spitzer\ IRS spectra of a large sample of protoplanetary disks to develop a diagnostic relation for the Humphreys $\alpha$ (7-6) line at 12.37 $\mu$m. This was the strongest \hi\ line in the IRS bandpass, but it is heavily contaminated by water emission and requires a correction that introduces large uncertainty when performed at the IRS resolution of $R\sim700$ (see Section \ref{acc_relations}). 

After the launch of \jwst, studies of MIRI-MRS disk spectra have so far adopted the \hi\ (7-6) line and the relation from \citet{rigliaco2015} in cases where water emission is not detected or very weak, implying that this line is free from contamination \citep{Beutheretal2023,Arulanantham2024,Tychoniecetal2024,vlasblom24}. The higher resolution of MRS, however, has demonstrated that this line can still include contamination from the nearby 11-8 line \citep[][and Section \ref{acc_relations} below]{Franceschietal2024,banzatti24}, suggesting that the relation by \citet{rigliaco2015} needs to be updated for use with MIRI-MRS spectra (see Section \ref{acc_relations}).

In this work we establish new relations for three selected \hi\ lines in the MIRI-MRS bandpass, by correlating their line luminosities to near-simultaneous, ground-based accretion luminosity measurements. We focus on two uncontaminated lines, \hi\ (10-7) at 8.76 $\mu$m and \hi\ (8-7) at 19.07 $\mu$m, as well as the water-contaminated \hi\ (7-6) line (12.37 $\mu$m) introduced by \citet{rigliaco2015}, for which we develop a new method to correct for the common case of contamination from water emission. 

This first paper is part of a large, multi-observatory effort coordinating five different instruments from space and ground that monitored a Class II spectroscopic binary that exhibits predictable accretion bursts. A summary of our global monitoring effort and the observed changes in disk chemisty will be described in an upcoming paper (Hyden et al.\ in prep.). In this first paper we include the datasets from \jwst\ MIRI-MRS complemented with the Las Cumbres Observatories 1m network (LCO) and VLT--X-Shooter (see Section \ref{sec: data}), which provided a densely sampled measure of the accretion luminosity to calibrate the MIRI data. 

The source, DQ Tau, is the binary pulsed-accretion archetype, exhibiting accretion bursts every orbital period ($P\sim 15.8$ d) as the system undergoes periastron passage \citep{Basrietal1997,Mathieuetal1997,Tofflemireetal2017a,Muzerolleetal2019,fiorellino2022}. This behavior results from dynamical resonances driven by the eccentric binary orbit ($e=0.57$) that clear the central region of the disk and spur periodic accretion streams that temporarily bridge the circumbinary disk and the central stars, which in turn, fuel periastron accretion bursts \citep[e.g.,][]{Artymowicz&Lubow1996,Munoz&Lai2016}. The system provides an ideal laboratory to, essentially, schedule observations over a large range of predictable accretion rates within a single observing season. Most importantly, monitoring variability in a single system removes significant sources of uncertainty and scatter that are present when developing diagnostic relations from a stellar sample where multiple properties can vary (distance, extinction, irradiation, environment, orientation, etc.). 

Figure \ref{fig: lightcurve_epochs} presents an overview of the three datasets included in this first paper, highlighting the time-series accretion luminosity measurements from LCO (purple circle) and X-Shooter (red diamonds), with respect to our \jwst\ epochs (vertical blue lines). Dotted vertical lines mark the binary periastron passages, where accretion bursts arrive like clockwork. In the lower panels, MIRI-MRS spectra centered on the \hi\ (10-7) line visually trace the changes in the accretion luminosity. 

In the closing of this paper, we compare our relations to archival MIRI-MRS observations from the JDISCS \citep[][Arulanantham et al.\ submitted]{pontoppidan24} and MINDS \citep{MINDS24} surveys using literature accretion luminosities (non-simultaneous), and finally, we compute accretion luminosities directly for each disk from the observed spectra. The accurate relations derived here, enabled by this exceptional near-simultaneous dataset, will provide a fundamental tool for disk studies with \jwst.

\section{Observations \& Data Reduction} \label{sec: data}

\subsection{\jwst--MIRI}
\label{jwst_obs}
The infrared spectra analyzed in this work were taken with the James Webb Space Telescope \citep[\jwst,][]{Gardner23} between September 11, 2024 and October 19, 2024 as part of Cycle 3 GO program 4727 (PI: A. Banzatti, Co-PIs: B. Tofflemire, C. Rab). DQ~Tau was observed at 4.9--28\,$\mu$m with the Medium Resolution Spectrometer \citep[MRS,][]{jwst-mrs,Argyriou23} mode on the Mid-Infrared Instrument \citep[MIRI,][]{miri,miri2}. The spectra were extracted and wavelength-calibrated with the JDISCS pipeline as described in \cite{pontoppidan24}, which adopts the standard MRS pipeline \citep{MIRI_pip} up to stage 2b and then uses observed asteroid spectra from GO program 1549 as calibrators to provide high-quality fringe removal and characterization of the spectral response function to maximize S/N in channels 2--4. A standard star is used in channel 1 where asteroid spectra have low S/N. To ensure similar fringes and maximize the quality of their removal, target acquisition with the MIRI imager was adopted to reach sub-spaxel precision in placing science target and asteroids on the same spot on the detector. For the data included in this work, we used MRS pipeline version 11.17.19 and Calibration Reference Data System context jwst\_1253.pmap. The MIRI spectra were continuum-subtracted using median smoothing and a 2nd-order Savitzky-Golay filter with the procedure presented in \cite{pontoppidan24}, updated to apply a final offset based on line-free regions as demonstrated in \cite{banzatti24}. (The data described above can be found in MAST: \dataset[10.17909/2sqp-pd37]{http://dx.doi.org/10.17909/2sqp-pd37}.)

\subsection{VLT--X-Shooter}
\label{xshooter_obs}

We observed DQ Tau with the X-Shooter spectrograph \citep{vernet11} mounted on the ESO Very Large Telescope (VLT) during the Service Mode program 114.2799.001 (PIs: C.F. Manara, B. Tofflemire). This spectrograph works at medium resolution ($R\sim$10,000 -- 20,000) and simultaneously covers the wavelength range $\sim$300-2500 nm, dividing the spectra in three arms. Our observations are set to achieve the highest possible resolution in the three arms, therefore using the narrowest slits (0.5", 0.4", 0.4" in the three arms), while also achieving absolute flux calibration accounting for the slit losses with a short exposure with the broad slits of 5.0" width prior to the narrow slit observations. A nodding cycle consisting of four positions ABBA was used for the narrow slit observations to perform sky subtraction. 

For each \jwst--MIRI observations, we requested to obtain three spectra with X-Shooter, for a total of 27 epochs. The observations were time constrained to match the \jwst--MIRI observations as close in time as possible. An absolute time interval was set to start the X-Shooter monitoring in each accretion burst covered by \jwst--MIRI, and the subsequent epochs were put in relative time-link intervals with minimum distances between the observations of 10 hours. 
Not all of the planned 27 spectra were taken due to different causes, including interventions on the telescopes or visitor mode runs. Missed epochs were rescheduled after the JWST program, but with contemporaneous LCO photometry. Data were taken between September 6th, 2024 and January 3rd, 2025 (UTC), always with excellent sky transparency conditions (``clear'' or ``photometric''). In this paper the 25 spectra taken during the aforementioned period are used in our analysis (Table \ref{tab: xshooter}).

Data were reduced using the ESO pipeline for X-Shooter v.3.6.8 \citep{modigliani2010} in the Reflex environment \citep{reflex}. Telluric lines were removed with the ESO Molecfit tool \citep{molecfit}, and the final flux calibration was obtained using the wide slit spectra, following the procedure described by \citet{manara2021}. 

\clearpage

\subsection{LCO--1m Network}
\label{lco_obs}

We monitored DQ Tau with multi-color time-series photometry using the Las Cumbres Observatory (LCO) 1 meter Telescope Network \citep{Brownetal2013}. The campaign covered more than 10 of the binary's orbital periods from August 13th, 2024 to January 22nd, 2025 (UTC; 162 days). These data provide a densely-sampled backdrop of the system's variability and enable a more complete interpretation of our program's MIRI and X-Shooter observations. 

LCO's flexible queue-scheduled observing allowed for a multi-component campaign. A moderate cadence program obtained visits every 12-18 hours in a full suite of filters (SDSS \up, {\it g}$^\prime$, {\it r}$^\prime$, {\it i}$^\prime$, and PANSTARRS {\it Z}, {\it Y}). An additional high-cadence program sought hourly visits in the 48 hours prior to a \jwst-MIRI observation in a subset of filters (SDSS \up, {\it r}$^\prime$, and PANSTARRS {\it Z}). In each case, a visit consists of three exposures per filter to increase the dynamic range of the observation. The full multi-color data set will be presented in a future work. Here, we focus on the SDSS \up, and {\it r}$^\prime$ results which had individual exposure times of 180 s and 3 s, respectively. 

Twelve 1-meter telescopes across five sites collected data for our campaign, each outfitted with a Sinistro 4096$\times$4096 CCD (0$\farcs$39 pixel$^{-1}$). Standard image calibrations were performed by the LCO facility pipeline (BANZAI; \citealt{McCullyetal2018}). To derive a differential magnitude time-series from this heterogeneous data set (multiple telescope systems, diverse observing conditions,  etc.), we employ an ensemble photometry approach \citep[e.g.][]{Honeycutt1992}. A detailed description of the methodology used here can be found in \citet{Tofflemireetal2017a}. In short, a linear regression is performed to minimize the variability of a suite of calibrating stars in the field of view. Weights assigned to each star in each image allow all non-variable stars to contribute to the solution when they are detected. This feature is ideal for our data set as pointing, sensitivity, and sky background (from the Moon, for instance) are all highly variable. The median uncertainty of the resultant light curve is 0.03 mag and 0.02 mag in the \up\ and \rp\ filters, respectively.

\begin{figure}
\centering
\includegraphics[width=0.48\textwidth]{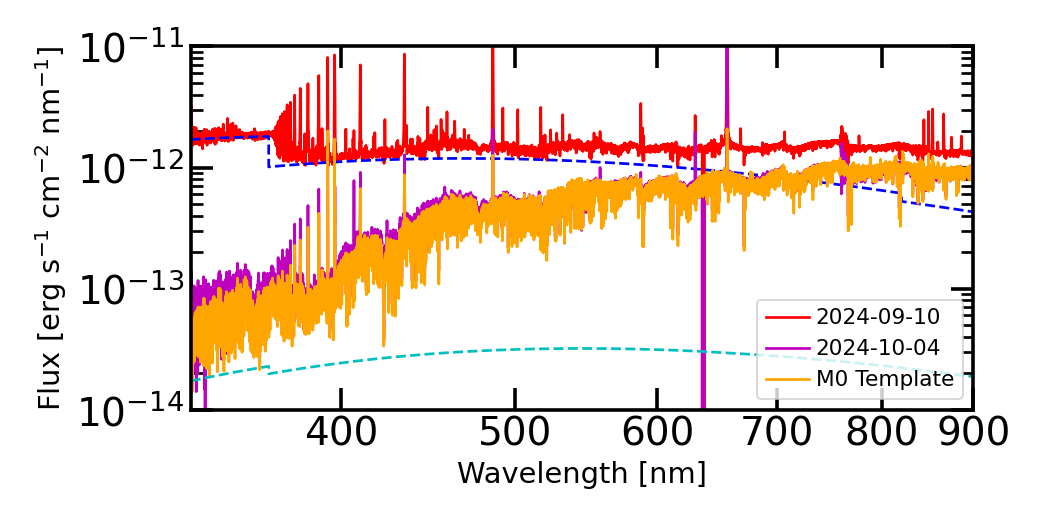}
\caption{X-Shooter spectra of DQTau in high and low accretion phases observed in this monitoring. The reddening corrected spectra (red and magenta) are fit with a combination of a photospheric template (orange) and a slab model (blue and cyan dashed lines). The accretion luminosity values measured from X-Shooter spectra are used to calibrate the LCO photometry (Figure \ref{fig: lacc_relation}).}
\label{fig: xs_fit_examples}
\end{figure}

\section{Ground-Based Accretion Luminosity Measurements}

\subsection{X-Shooter}

The X-Shooter spectra allow us to measure the accretion luminosity by fitting the Balmer Jump and the broad-band spectrum with a combination of a photospheric template and a model for the emission of the accretion shock, plus a characterization of the interstellar extinction. Following \citet{Manaraetal2013fitter}, the set of photospheric templates is chosen from a collection of non-accreting young stars \citep{Manaraetal2013,Manaraetal2017,Claesetal2024} and a slab model is used to reproduce the continuum emission due to accretion. Finally, the extinction law is taken from \citet{Cardellietal1989} using a value of $R_V=3.1$. In this work the assumption is made that the combined spectrum of the two equal-mass components of the binary can be treated as a single star.

The spectra are fit in a number of features between the Balmer continuum at $\lambda\sim330$ nm and $\lambda\sim750$ nm, and the best fit is found by minimizing a $\chi^2_{\rm like}$ function. In all cases, the best fit is found with a template of spectral type M0, and extinction values of $A_V\sim1.6\pm0.2$ mag, with only a few exceptions where an $A_V$ of $\sim2$ mag was preferred by the fit. From the parameters of the best-fit template we derive a stellar luminosity ($L_\star$) of $1.5\pm0.2 L_\odot$. Uncertainties in the above measurements represent the deviation in the fit values across the observations. 

The stellar masses ($M_\star$) are derived assuming stellar parameters from the \citet{Baraffeetal2015} evolutionary model interpolation on the HR Diagram, obtaining $M_\star\sim0.60\pm0.03 M_\odot$. This formal error does not include systematic model-dependent uncertainties that are typically $\sim0.1 M_\odot$. The stellar masses and their uncertainties only factor into the mass accretion rate, and do not affect the primary focus of our study, the accretion luminosity. These values are, however, in good agreement with the study by \citet[$M_1=0.63\pm0.13M_\odot$, $M_2=0.59\pm0.13M_\odot$]{Czekalaetal2016}, that performed a joint fit of the stellar radial velocities and the circumbinary CO rotation curve.

By measuring the flux of the slab model, we infer the accretion luminosity (\lacc) with a typical relative uncertainty of $\sim$0.1 dex (see Section \ref{lco_lacc}), and the mass accretion rate ($\dot{M}_{\rm acc}$). These two parameters change dramatically from one epoch to another, as expected for this variable system. Examples of best fits at a high and low level of accretion are shown in Fig.~\ref{fig: xs_fit_examples}.
The values are reported in Table \ref{tab: xshooter}, along with the absolute time offset from the nearest LCO visit ($\Delta t$). In all spectra the accretion luminosity is measured to be well above the X-Shooter \lacc\ noise floor ($\rm{log}_{10}($\lacc$/L_\odot)\sim-2$), which is empirically determined from a population of non-accreting T Tauri-stars \citep{Manaraetal2013,Claesetal2024}.

\begin{deluxetable}{l c c c c}
\tablecaption{\label{tab: xshooter} DQ Tau Measurements with X-Shooter}
\tablehead{
  \colhead{Date} & 
  \colhead{$\rm{log}_{10}(L_{acc}/L_\odot)$} & 
  \colhead{$\rm{log}_{10} \dot{M}$} & 
  \colhead{$A_V$} &
  \colhead{LCO $\Delta t$} \\
  \colhead{(BJD)} & 
  \colhead{} & 
  \colhead{($M_\odot$/yr)} & 
  \colhead{$A_V$} &
  \colhead{(hr)}
}
\startdata
2460559.86 & -1.61 & -8.38 & 1.5 & 3.01 \\
2460560.88 & -1.21 & -7.98 & 1.6 & 0.6 \\
2460562.84 & -0.26 & -6.95 & 2.0 & 0.55 \\
2460563.82 & 0.03 & -6.72 & 1.7 & 2.07 \\
2460564.84 & 0.14 & -6.55 & 2.1 & 0.11 \\
2460571.85 & -0.93 & -7.69 & 1.6 & 2.79 \\
2460572.81 & -0.99 & -7.75 & 1.5 & 1.56 \\
2460573.78 & -1.05 & -7.82 & 1.5 & 11.93 \\
2460583.85 & -1.27 & -8.02 & 1.6 & 0.13 \\
2460585.84 & -1.15 & -7.92 & 1.5 & 5.16 \\
2460586.81 & -1.28 & -8.04 & 1.5 & 3.01 \\
2460587.79 & -1.54 & -8.31 & 1.5 & 0.35 \\
2460595.75 & -0.37 & -7.06 & 1.9 & 3.85 \\
2460598.75 & -1.29 & -8.05 & 1.5 & 2.88 \\
2460599.74 & -1.25 & -7.97 & 1.6 & 0.69 \\
2460600.72 & -1.13 & -7.9 & 1.5 & 6.55 \\
2460605.73 & -1.33 & -8.09 & 1.5 & 1.38 \\
2460634.84 & -1.31 & -8.07 & 1.5 & 0.86 \\
2460635.63 & -1.23 & -7.99 & 1.5 & 0.85 \\
2460647.60 & -1.23 & -7.99 & 1.5 & 1.06 \\
2460649.67 & -0.73 & -7.44 & 1.6 & 1.45 \\
2460651.59 & -0.71 & -7.44 & 1.6 & 1.11 \\
2460671.58 & -1.37 & -8.13 & 1.5 & 15.21 \\
2460673.59 & -0.73 & -7.44 & 1.5 & 4.39 \\
2460678.60 & -1.14 & -7.78 & 1.9 & 2.40 \\
\enddata
\end{deluxetable}

\begin{figure}
\centering
\includegraphics[width=0.47\textwidth]{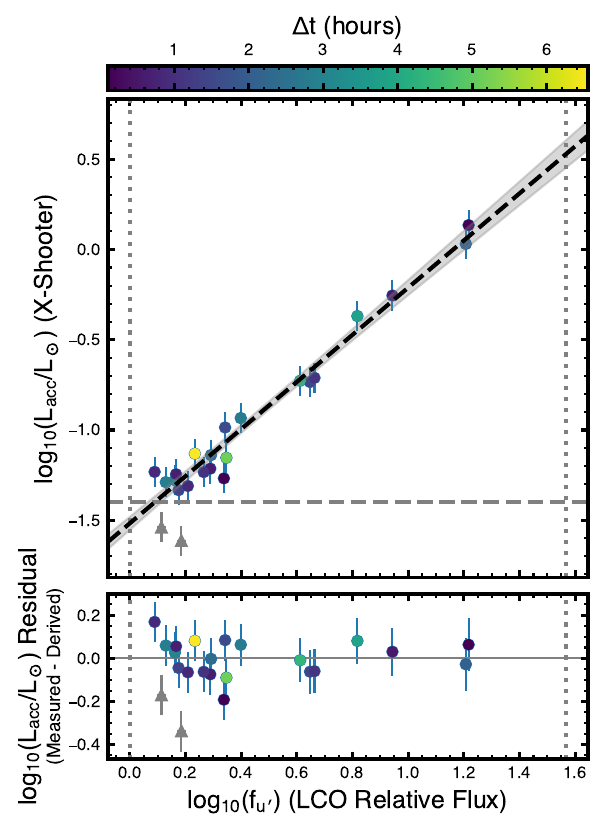}
\caption{Correlation between the accretion luminosity measured from X-Shooter spectra and \up\ relative flux measured from LCO photometry. All points represent near-simultaneous measurements from both instruments ($\Delta t \lesssim 5$ hr). The dotted line is a linear fit. We adopt log$_{10}$(\lacc/\lsun) $= -1.4$ as the LCO measurement floor, signified by the horizontal dashed line. Accretion luminosity measurement below this value are not included in the fit (grey triangles). Vertical dotted lines signify the range of relative flux measurements across the program. The bottom panel presents the residual between the X-Shooter measured accretion luminosity and the value derived from the \up\ relative flux. The tight correlation and modest residuals (residual RMS = 0.12 dex) allow for a robust derivation of the accretion luminosity from the time-series photometry.}
\label{fig: lacc_relation}
\end{figure}

\subsection{LCO}
\label{lco_lacc}

Deriving accretion luminosities from near-UV or blue-optical photometry is an established practice \citep[e.g.,][]{Gullbringetal1998,Robbertoetal2004,Manaraetal2012,Venutietal2014}. The approach typically requires absolute photometry, measuring the excess bandpass luminosity above a model for the stellar photosphere. In this work, we rely on differential photometry alone by bootstrapping off precise accretion luminosity measurements from near-simultaneous X-Shooter observations. We derive a relation between the relative \up-flux and the X-Shooter accretion luminosity to compute accretion luminosities across the densely-sampled LCO time-series. 

Figure \ref{fig: lacc_relation} presents the relation between the \up\ relative flux and the near-simultaneous X-Shooter \lacc. The data points are color-coded by the time lag ($\Delta t$) between observations; we only consider measurements made within the same night ($\Delta t < 8$ hr) as to avoid additional scatter from non-simultaneous observations. Below a certain accretion luminosity, log$_{10} ($\lacc$/L_\odot) \sim -1.4$, the \up\ flux no longer scales linearly. The \up\ flux levels out while the more sensitive X-Shooter measurements probe lower \lacc\ values. We set this value as the \lacc\ measurement floor for the \up\ photometry and designate the value in Figures \ref{fig: lightcurve_epochs} and Figure \ref{fig: lacc_relation} as a gray horizontal line. The X-Shooter measurements below this value (and with $\Delta t < 8$ hr) are included in Figure \ref{fig: lacc_relation} as gray triangles for completeness, but are not used to derive the scaling relation. We note that this is not a limitation of photometric detection, but rather where the accretion luminosity approaches a level that is comparable to the light curve noise, both from the pipeline precision and the variability intrinsic to the stellar photosphere(s).

We use the {\tt linmix} package \citep{Kelly2007} to perform the regression in a Bayesian framework, incorporating uncertainties from both measurements. The relative X-Shooter uncertainties are estimated to be 0.08 dex based on an initial Monte-Carlo exercise by measuring the average deviation from least-squares fits using random sampling with replacement. On an absolute scale, a typical uncertainty of 0.2 dex is assumed from previous X-Shooter studies \citep[e.g.][]{Manaraetal2017}. The best fit is shown as the dashed black line in Figure \ref{fig: lacc_relation}, with the corresponding 1$\sigma$ confidence interval shown in gray. Residuals between the measured and derived \lacc\ are presented in the bottom panel, which have an RMS of 0.08 dex. Since the relative flux is arbitrarily scaled, the fit parameters are not applicable to other data sets, but we note the slope of the relation is $1.31\pm0.07$.

This relation is tighter than those derived from samples of accreting stars. While ours is a relative relation tailored to DQ Tau, rather than absolute relation, the derived LCO accretion luminosities are robust because we consider a single source (i.e., constant distance, extinction, stellar parameters) with near-simultaneous photometry and spectroscopy, which bypass many of the compounding sources of error in compiled samples. 

The result of applying this relation to the full LCO time-series is presented in Figure \ref{fig: lightcurve_epochs}. We also present the derived mass accretion rate (\mdot\ in $M_\odot$ yr$^{-1}$) on the right-hand y-axis. This transformation is made by computing a linear relation between the X-Shooter \lacc\ and \mdot\ measurements and applying it to the derived LCO accretion luminosities.

\subsubsection{A Reddening Event}
\label{reddening}

We detect a reddening event during MIRI Epoch 3 (Sept 26th, 2024; JD $\sim$ 2460580), where the \up$-$\rp\ color exceeds the local color-magnitude relation defined by our data by more than 3$\sigma$. The photometric measurements are robust -- three LCO visits made within 1.6 hours from two different observatory sites confirm the event. Additionally, the visit $\sim$7 hours prior to the aforementioned trio is also exceptionally red (2.7$\sigma$) and is likely part of the same reddening event. The outlying red color and corresponding dimming in the light curve (Figure \ref{fig: lightcurve_epochs}), are suggestive of ``dipper''-like behavior, which is seen in a subset of accreting sources \citep[e.g.,][]{Codyetal2014}. Dipper events are thought to result from time-variable dust obscuration from the inner disk and/or accretion streams \citep{Bouvieretal2007,Ansdelletal2016a}. 

With evidence for a dust obscuration event, the LCO-derived accretion luminosity is a lower limit on the true value. Correcting the observed flux measurements requires modeling the size distribution of the obscuring dust grains, which likely differs from the interstellar medium \citep[e.g.,][]{Sitkoetal2023}, and is beyond the scope of the current work. Given the intrinsic value of this high-accretion luminosity epoch, we estimate the accretion luminosity at the time of MIRI Epoch 3 using the observed variability in the DQ Tau light curve. We leverage the known flux difference ($\delta f_{gap}$) on either side of the reddening event's duration ($\delta t_{gap}$) to characterize the intermediate flux values DQ Tau exhibits during regions of the light curve with similar behavior. The full gapped duration is $\delta t_{gap}=1.67$ d and the time lag to MIRI Epoch 3 is $\delta t_{MIRI, 3} = 0.96$ d. This region is highlighted in Figure \ref{fig: lightcurve_epochs} with a vertical gray band. In detail, we densely over sample the relative flux light curve with a linear interpolation and identify regions that have $\delta f$ values across $\delta t_{gap}$ that are within $3\sigma$ of $\delta f_{gap}$ (i.e., the value that spans MIRI Epoch 3). We find 35 discrete regions that satisfy this condition. For points in each region, we compile the flux difference at the $\delta t_{MIRI,3}$ time lag. Adding this $\delta f$ distribution to the flux value at the start of the reddening event gap, we arrive at a distribution of values at the time of MIRI Epoch 3 informed by the light curve's behavior. We adopt the median of this distribution as our estimate and, conservatively, the 99.8\% interval ($\sim3\sigma$) as our uncertainty. The corresponding \lacc\ uncertainty is 4-10 times larger than our typical measurement error. The result is shown in Figure \ref{fig: lightcurve_epochs} as a black point and the value is listed in Table \ref{tab: accretion values} for MIRI Epoch 3.

\begin{figure*}
\centering
\includegraphics[width=1\textwidth]{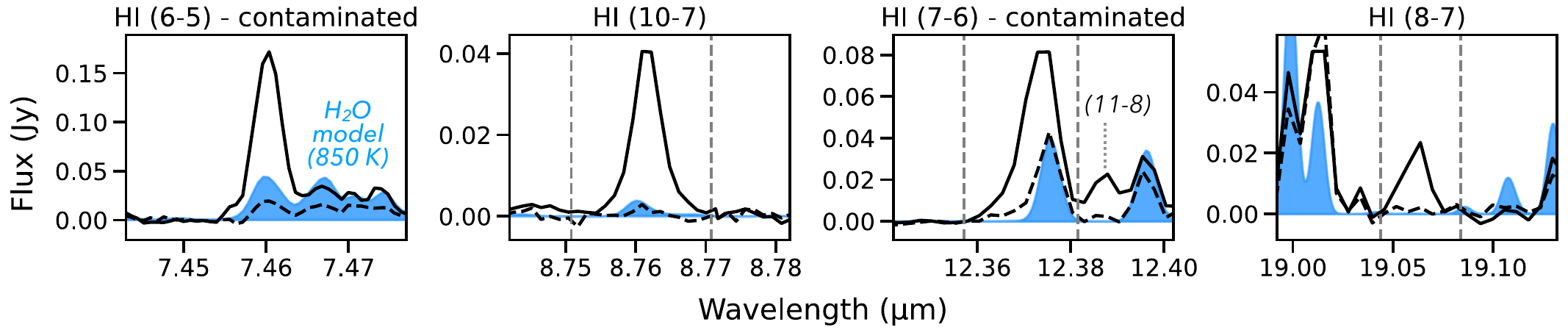}
\caption{Gallery of some strong hydrogen recombination lines observed in MIRI spectra. Two DQ~Tau epochs are shown in each panel to illustrate a high accretion phase (September 11th, in solid black) and the lowest accretion phase in this program (October 7th, dashed black). A hot water model (850~K) is shown in blue for reference. The (10-7) and (8-7) lines are selected from those identified in \cite{banzatti24} as free of contamination from molecular emission. Two examples of significant contamination from water are shown for reference: the (7-6) line \citep[used in][]{rigliaco2015} and the (6-5) line, the strongest \hi\ line observable with MIRI-MRS, but contaminated with ro-vibrational water emission. The vertical gray dashed lines indicate the ranges where the line flux is measured, in the case of the (7-6) line to minimize contamination from the nearby (11-8) line. }
\label{fig: HI_series}
\end{figure*}

\begin{deluxetable*}{l c c c c c c c c c}
\tabletypesize{\small}
\tablewidth{0pt}
\tablecaption{\label{tab: accretion values} Line and accretion luminosities measured for DQ~Tau in this program.}
\tablecolumns{10}
\tablehead{
  \colhead{ } & 
  \multicolumn{3}{c}{$\rm{log}_{10}(L_{line}/L_\odot)$} &
  \colhead{} &
  \multicolumn{5}{c}{$\rm{log}_{10}(L_{acc}/L_\odot)$} \\
  \cline{2-4}
  \cline{6-10}
  \colhead{Epoch} & 
  \colhead{\hi\ (10-7)} & 
  \colhead{\hi\ (7-6)} & 
  \colhead{\hi\ (8-7)} & 
  \colhead{} &
  \colhead{LCO} & 
  \colhead{\hi\ (10-7)} & 
  \colhead{\hi\ (7-6)}& 
  \colhead{\hi\ (8-7)} & 
  \colhead{Average}
}
\startdata
1: Sep 11 & $-5.00(0.02)$ & $-4.85(0.04)$ & $-5.46(0.06)$ &  & $0.05(0.09)$ & $-0.18(0.17)$ & $-0.16(0.14)$ & $-0.23(0.35)$ &$-0.17(0.10)$ \\
2: Sep 14 & $-5.56(0.05)$ & $-5.19(0.08)$ & $-5.81(0.17)$ &  & $-1.17(0.05)$ & $-0.93(0.14)$ & $-0.83(0.17)$ & $-0.90(0.46)$ &$-0.89(0.10)$ \\
3: Sep 26 & $-4.98(0.02)$ & $-4.75(0.04)$ & $-5.30(0.04)$ &  & $-0.20(0.39)^{\dagger}$ & $-0.15(0.18)$ & $0.05(0.17)$ & $0.07(0.45)$ &$-0.03(0.12)$ \\
4: Oct 1 & $-5.85(0.05)$ & $-5.42(0.11)$ & $<-5.98$ &  & $-1.11(0.05)$ & $-1.33(0.19)$ & $-1.29(0.24)$ & $<-1.22$ &$-1.31(0.15)$ \\
5: Oct 5 & $<-5.87$ & $-5.52(0.12)$ & $<-5.88$ &  & $-1.33(0.04)$ & $<-1.36$ & $-1.49(0.25)$ & $<-1.04$ &$-1.49(0.25)$ \\
6: Oct 7 & $<-5.85$ & $-5.58(0.09)$ & $<-5.96$ &  & $-1.36(0.04)$ & $<-1.33$ & $-1.62(0.22)$ & $<-1.19$ &$-1.62(0.22)$ \\
7: Oct 11 & $-5.08(0.02)$ & $-4.99(0.03)$ & $-5.61(0.08)$ &  & $-0.40(0.07)$ & $-0.29(0.15)$ & $-0.43(0.11)$ & $-0.52(0.31)$ &$-0.39(0.09)$ \\
8: Oct 13 & $<-5.87$ & $-5.33(0.07)$ & $<-5.95$ &  & $-1.25(0.04)$ & $<-1.35$ & $-1.11(0.15)$ & $<-1.18$ &$-1.11(0.15)$ \\
9: Oct 19 & $-5.75(0.04)$ & $-5.39(0.09)$ & $-6.13(0.16)$ &  & $-1.24(0.04)$ & $-1.20(0.17)$ & $-1.24(0.20)$ & $-1.51(0.69)$ &$-1.22(0.13)$ \\
\enddata
\tablecomments{The \hi\ (7-6) line luminosity has been corrected for water contamination as described in Section \ref{acc_relations}. \\
$\dagger$ -- Measurement is affected by a reddening event, value and uncertainty determination described in Section \ref{reddening}.}
\end{deluxetable*}

\begin{figure*}
\centering
\includegraphics[width=1\textwidth]{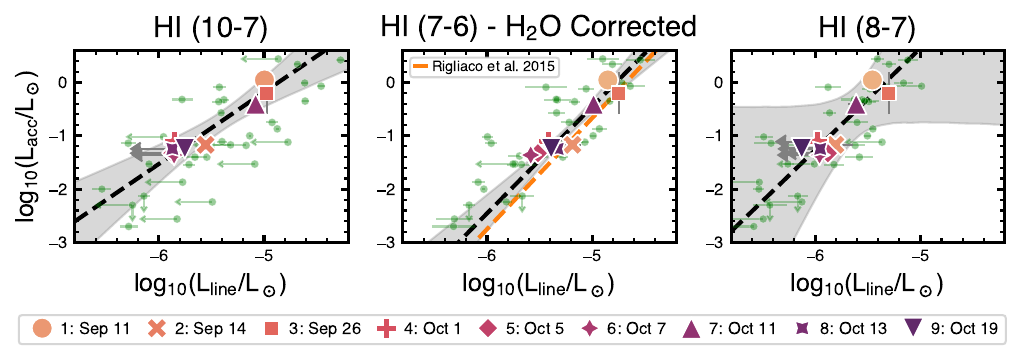}
\caption{Correlations between \hi\ line luminosities and \lacc\ as measured in this monitoring program. The \hi\ (7-6) luminosity has been corrected for water contamination as explained in Section \ref{acc_relations}. Large colored data points are from the DQ Tau epochs in this work. Smaller green circles show JDISCS and MINDS published spectra with non-contemporaneous literature \lacc\ measurements (see text). The larger scatter is likely due to non-synchronous measurements and varied \lacc\ measurement methodologies. For non detections, the 2-$\sigma$ upper limit is shown with an arrow. The black dashed lines show the median correlations as measured for DQ Tau, while the orange line in the middle panel shows the \hi\ (7-6) relation previously measured by \cite{rigliaco2015}. The offset with the new relation found from MIRI is due to the limited correction from contamination that was possible with IRS (see Section \ref{acc_relations}).}
\label{fig: HI_Lacc_correlations}
\end{figure*}

\section{Calibration of MIRI accretion relations}
\label{acc_relations}
For each MIRI epoch, we use LMFIT \citep{lmfit} within iSLAT \citep{iSLAT,iSLAT_code} to fit and measure two \hi\ lines identified in \cite{banzatti24} as relatively strong and uncontaminated by molecular emission, especially from water (Figure \ref{fig: HI_series} and Table \ref{tab: accretion values}): (10-7) at 8.76~$\mu$m and (8-7) at 19.06~$\mu$m. The variability in the \hi\ (10-7) line over our campaign is shown in Figure \ref{fig: lightcurve_epochs}. We identify this line as the best accretion tracer due to the combination of its relative strength and minimal contamination by water. 

Other \hi\ lines covered by MIRI that could potentially be used for this work are excluded as being weaker (12-8; 10.50~$\mu$m) or more significantly contaminated by water (9-6; 5.91~$\mu$m) or OH \citep[9-7; 11.3~$\mu$m; see Figures 1--4 in][]{banzatti24}. Additionally, \hi\ lines at $<8$~$\mu$m have been observed in absorption in one disk \citep{salyk25} as well as in our low-accretion DQ~Tau epochs (see Appendix \ref{app: HI_abs}), making them unsuitable for accretion measurements. The strongest \hi\ line available at MIRI wavelengths, the (6-5) line at 7.46~$\mu$m, is contaminated by ro-vibrational water transitions in the $v=1-0$ band (Figure \ref{fig: HI_series}). Although contamination from water and other species can, in principle, be removed by fitting and subtracting a model, the uncertainty introduced by this process is currently unknown in the case of ro-vibrational emission lines that are affected by non-LTE excitation \citep{meijerink09,banzatti23,banzatti24}. In this work, we wish to provide accretion relations that can be generally applied to MIRI spectra without having to perform fits to molecular spectra first. The only contaminated line we include in our analysis is the \hi\ (7-6) line at 12.37~$\mu$m, introduced as accretion tracer in \cite{rigliaco2015}, which can be more easily corrected from purely rotational water emission at the MIRI-MRS resolution as explained below.

To test and compare with previous results from \cite{rigliaco2015}, we develop a new procedure to correct the \hi\ (7-6) line from water contamination as explained in Appendix \ref{app: correction}. We find that at the low resolving power of IRS, \cite{rigliaco2015} could not identify the correct water transition that is contaminating the \hi\ (7-6) line (see their Table 2): they used the nearby line centered at 12.396~$\mu$m, which MIRI spectra now clearly show is offset from the \hi\ (7-6) line (see Figure \ref{fig: HI_series}). Moreover, the \hi\ (11-8) line also contaminates the \hi\ (7-6) line at the IRS resolution, while MIRI clearly resolves them (Figure \ref{fig: HI_series}). Therefore, we can expect that the relation derived in \cite{rigliaco2015} may have an offset from the true accretion relation due to an over-estimation of the \hi\ (7-6) line flux, which we indeed find below. The new method we develop to correct for water contamination generally adds only a $\lesssim 10\%$ uncertainty to the measured \hi\ (7-6) line flux, as demonstrated in Appendix \ref{app: correction}. This method is easier than performing a model fit to the rotational water emission and can be easily applied to large disk samples. Moreover, at the resolving power of MRS it is possible to minimize the contamination from the nearby \hi\ (11-8) line by measuring the \hi\ (7-6) line flux over the range indicated in Table \ref{tab: lin_fit_params} and Figure \ref{fig: HI_series}. The water corrected line luminosity for \hi\ (7-6), and the uncontaminated (10-7) and (8-7) line luminosities are presented in Table \ref{tab: accretion values}.

Figure \ref{fig: HI_Lacc_correlations} shows the strong correlations observed between the \hi\ line luminosities from MIRI and \lacc\ measurements from LCO and X-shooter. We use the \lacc\ values from LCO data that were taken closest in time to each MIRI epoch, as reported in Table \ref{tab: accretion values} (time lags are typically less than a few hours, see Figure \ref{fig: lightcurve_epochs}). The correlations have Pearson coefficients above 0.9 and small scatter indicative of the quality of this exceptional, near-simultaneous monitoring program across three telescopes. The relationship with the (8-7) line is, however, affected by higher noise in MIRI channel 4.

We fit linear relations to these data in a Bayesian framework using the {\tt linmix} package. Each fit has the form: 
\begin{equation}
    \rm{log}_{10}(L_{acc}/L_\odot) = a \times [(\rm{log}_{10}(L_{HI}/L_\odot) - \rm{Shift}] + b
    \label{eqn: fit}
\end{equation}
where a constant shift is introduced to minimize the covariance between the slope ($a$) and the y-intercept ($b$). Table \ref{tab: lin_fit_params} includes the linear regression parameters for all three correlations.

The correlation measured in the \hi\ (7-6) line, after correction for water contamination, shows an offset in comparison to the result reported previously from \spitzer\ data by \citet{rigliaco2015}, but a consistent slope ($2.08 \pm 0.39$; orange line in Figure \ref{fig: HI_Lacc_correlations}). As noted above, this offset is the consequence of the IRS resolution, which resulted in unresolved contaminating flux. The higher resolving power of MIRI spectra now provides an improvement on previous results obtained from the IRS.

We use the accretion relations provided in Table \ref{tab: lin_fit_params} to measure \lacc\ from each of the three \hi\ lines for each epoch of DQ~Tau MIRI spectra. The results are reported in Table \ref{tab: accretion values}, where we also include the weighted average accretion luminosity obtained from the three \lacc\ values. We note that the \hi-derived \lacc\ value for epoch 3 is insensitive to the reddening event discussed above in Section \ref{reddening}, highlighting another case where mid-infrared relations may be helpful for accretion monitoring studies of young stars with dusty disks or envelopes.

In Figure \ref{fig: HI_Lacc_correlations}, we also include a comparison of all the MIRI disk spectra of low- and solar-mass stars currently published by the JDISCS and MINDS collaborations \citep[][Arulanantham et al.\ submitted]{grant23,grant24,perotti23,gasman23,gasman25,schwarz24,temmink24,MINDS24,pontoppidan24,banzatti23b,banzatti24,vlasblom24}, using non-simultaneous \lacc\ values compiled from the literature. All the spectra were homogeneously reduced with the JDISCS pipeline as described in \cite{pontoppidan24}, and then continuum-subtracted and processed as described above for DQ~Tau in this work.

\begin{deluxetable*}{l c c c c c c c}
\tablecaption{\label{tab: lin_fit_params} Linear regression parameters for correlations in Figure \ref{fig: HI_Lacc_correlations}.}
\tablehead{\hi\ line & $\lambda$ [$\mu$m] & Line range [$\mu$m] & $a (\sigma_a)$ & $b (\sigma_b)$ & Shift & Pearson $R$ & Residual RMS}
\startdata
(10-7) & $8.76006$ & 8.7508--8.7708   & $ 1.35 ( 0.32 )$ & $ -0.72 ( 0.11 )$ & $ -5.40 $ & $ 0.94 $ & $ 0.17 $ \\
(7-6) & $12.37192$ & 12.3571--12.3815 & $ 2.00 ( 0.27 )$ & $ -0.85 ( 0.06 )$ & $ -5.20 $ & $ 0.93 $ & $ 0.20 $ \\
(8-7) & $19.06189$ & 19.0437--19.0839 & $ 1.91 ( 1.09 )$ & $ -0.54 ( 0.27 )$ & $ -5.62 $ & $ 0.89 $ & $ 0.25 $ \\
\enddata
\tablecomments{Linear relations are in the form $\rm{log}_{10}(L_{acc}/L_\odot) = a \times [\rm{log}_{10}(L_{HI}/L_\odot) - \rm{Shift}] + b$. The line range in the third column indicates the spectral range where the line flux is measured for each line (also shown in Figure \ref{fig: HI_series}.)}
\end{deluxetable*}

\section{Discussion}
\label{discusssion}

DQ Tau is an eccentric, short-period binary with accretion mechanics that are distinct from single stars. This distinction does not preclude its utility in defining a relation between \hi\ line and accretion luminosities. DQ Tau's equal-mass central stars are unlikely to host stable circumstellar disks that feed accretion funnels at a disk-magnetosphere interface \citep{Tofflemireetal2017a}. The periastron separation ($\sim12$ \rstar; \citealt{Czekalaetal2016}) is small enough that the typical magnetospheric truncation radii assumed for single stars would essentially overlap. Indeed, evidence for magnetospheric collision events have been observed for DQ Tau (e.g., synchrotron and X-ray flares; \citealt{Salteretal2010,Getmanetal2023}). This difference will affect the details of the mass accretion rate derivation ($\dot{M}$), but should be immune to the accretion luminosity used to derive our diagnostic relations. Whether the circumbinary accretion streams impact the stars directly, or become entrained in the stellar magnetospheres prior to impact, the end result is still a column of material feeding a standing shock with continuum and line emission that can be well-fit with a slab model used for single stars, as seen in Figure \ref{fig: xs_fit_examples} and \citet{fiorellino2022}. Simply put, it is the intensity of accretion that is relevant, not the details of the process. Lastly, the agreement between the DQ Tau measurements and the archival sample, which is predominantly composed of single stars (89\%) and some wide binaries ($>0.1\arcsec$; 8\%), support the broader validity of the results derived here.

NIR \hi\ lines have long been used as accretion tracers, with Brackett $\gamma$ and Pfund $\beta$ being at the longest wavelengths (2.16 and 4.65~$\mu$m respectively). With high-resolution spectra obtained from ground-based instruments, these lines have shown large broadening velocities consistent with magnetospheric stellar accretion (150--350~km/s). The MIR \hi\ line introduced as accretion tracer by \cite{rigliaco2015}, however, was unresolved with \spitzer-IRS. With the improved resolving power of the MIRI-MRS, the line width can, for the first time, be measured at MIR wavelengths \citep{banzatti24}. We measure the de-convolved \hi\ (10-7) line FWHM to be 130--190 km/s in the DQ~Tau data (with broader values observed during the accretion bursts), consistent with the FWHM measured in optical \hi\ lines covered with X-shooter in this work, as well as with the range of FWHM values (140--300~km/s) measured from MIRI spectra in the larger disk sample included in Figure \ref{fig: HI_Lacc_correlations}. These results confirm that MIR \hi\ lines have velocity broadening consistent with an accretion-flow origin.

\begin{figure*}
\centering
\includegraphics[width=1\textwidth]{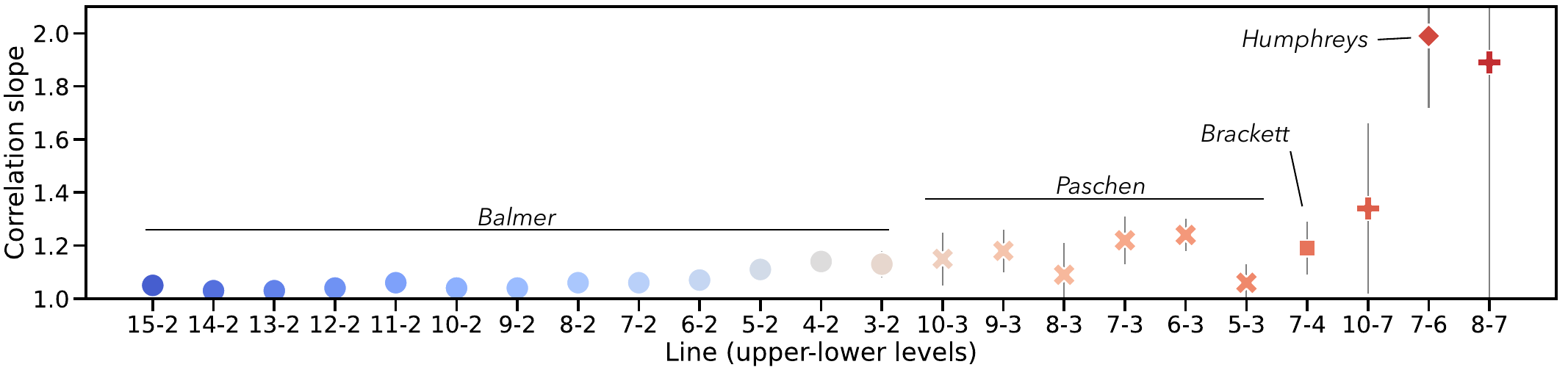}
\caption{Correlation slopes found in $\rm{log}_{10}($\lacc$/L_\odot)$ relations with \hi\ line luminosities for different transitions, ordered by wavelength. Balmer, Paschen, and Brackett series slopes are from \cite{Alcalaetal2017}. Longer wavelength transitions are from this work. The combination of all these results suggests a trend of increasing slope with wavelength, which was already visible in the results from \cite{Alcalaetal2017}.}
\label{fig: Slope_vs_wl}
\end{figure*}

In comparison with the slopes of accretion relations in the optical and NIR, we find our MIR relations are steeper. At shorter wavelengths, \citet{Alcalaetal2014,Alcalaetal2017} compiled a set of \hi\ line relations with slopes between 1.03 and 1.24, reporting an increasing trend in the slope with wavelength from 0.37 to 2.16 $\mu$m. This trend now appears to extend to the MIR, where we find slopes from 1.3 to 2, from 8.76 to 12.37 $\mu$m (Figure \ref{fig: Slope_vs_wl}). This trend might be due to the fact that the lines are probing different parts of the accretion column or post-shock region, and their relative relation with \lacc\, is related to the physical conditions (e.g., temperature, density, and optical depth) in that part of the accretion column. The different lines would have a different contribution to the total accretion luminosity depending on how much material along the accretion columns emits in those transitions. Future studies should follow-up on this observed trend by investigating the impact accretion column structure has on hydrogen emission line ratios \citep[e.g.,][]{KwanFischer}. 

One key result from this work is the fact that the relation between the \hi\ line and accretion luminosity, are tighter for the DQ Tau than the archival disk sample available from previous work (Figure \ref{fig: HI_Lacc_correlations}). This is readily explained by the fact that using a single target reduces the uncertainties introduced in accretion estimates from different extinctions, distances, system geometries, and stellar properties. Equally importantly, the simultaneity of the \lacc\ measurements in this monitoring program lowers the scatter introduced by accretion variability, known to change accretion rates in classical T-Tauri stars by a factor $\sim$3 on timescales as short as days \citep[e.g.,][]{Costiganetal2014,Zsidietal2022,FischeretalPPVII}. The non-simultaneous observations in the JDISCS and MINDS samples, shown in Figure \ref{fig: HI_Lacc_correlations}, are likely introducing scatter due to accretion variability between the date of the accretion measurement and that of the MIRI spectrum. The archival sample has residual RMS values of 0.52, 0.56, and 0.50 for the (10-7), (7-6), and (8-7) lines, respectively, compared to a factor of 2--3 smaller for DQ Tau.
Given the scientific value of having simultaneous accretion values, we measure the accretion luminosity for each disk in the archival sample using Equation \ref{eqn: fit} and the fit values in Table \ref{tab: lin_fit_params}. The line luminosities, derived \lacc, and the average \lacc\ value (error-weighted mean) across the three diagnostic lines are presented in Table \ref{tab: archival sample} in the same format used for DQ~Tau in Table \ref{tab: accretion values}.

\begin{deluxetable*}{l c c c c c c c c c c}
\tabletypesize{\footnotesize}
\tablewidth{0pt}
\tablecaption{\label{tab: archival sample} \hi\ Luminosities and Line-Derived Accretion Luminosities for Disks with archival MIRI-MRS data published to date}
\tablecolumns{11}
\tablehead{
  \colhead{} &
  \multicolumn{3}{c}{$\rm{log}_{10}(L_{line}/L_\odot)$} &
  \colhead{} &
  \multicolumn{4}{c}{$\rm{log}_{10}(L_{acc}/L_\odot)$} \\
  \cline{2-4}
  \cline{6-9}
  \colhead{Target} &
  \colhead{\hi\ (10-7)} & 
  \colhead{\hi\ (7-6)} & 
  \colhead{\hi\ (8-7)} & 
  \colhead{} &
  \colhead{\hi\ (10-7)} & 
  \colhead{\hi\ (7-6)}& 
  \colhead{\hi\ (8-7)} & 
  \colhead{Average} &
  \colhead{Program ID} &
  \colhead{Date}
}
\startdata
A S205 N & $-4.46(0.04)$ & $-4.64(0.14)$ & $<-4.99$ &  & $0.55(0.33)$ & $0.27(0.32)$ & $<0.66$ &$0.41(0.23)$ &1584&4-Apr-23\\
AS 209 & $<-4.77$ & $-5.07(0.04)$ & $<-5.37$ &  & $<0.12$ & $-0.60(0.10)$ & $<-0.05$ &$-0.60(0.10)$ &1584&28-Aug-23\\
BP Tau & $<-5.32$ & $-5.34(0.05)$ & $<-5.95$ &  & $<-0.61$ & $-1.13(0.12)$ & $<-1.17$ &$-1.13(0.12)$ &1282&20-Feb-23\\
CI Tau & $-4.87(0.01)$ & $-4.93(0.07)$ & $-5.45(0.03)$ &  & $-0.01(0.20)$ & $-0.32(0.17)$ & $-0.22(0.34)$ &$-0.20(0.12)$ &1640&27-Feb-23\\
CX Tau & $<-5.83$ & $-6.31(0.17)$ & $<-6.49$ &  & $<-1.30$ & $-3.07(0.46)$ & $<-2.21$ &$-3.07(0.46)$ &1282&20-Feb-23\\
DF Tau & $-4.92(0.05)$ & $-5.22(0.08)$ & $-5.60(0.10)$ &  & $-0.08(0.20)$ & $-0.89(0.17)$ & $-0.51(0.34)$ &$-0.55(0.12)$ &1282&20-Feb-23\\
DL Tau & $-4.61(0.03)$ & $-4.83(0.03)$ & $-5.53(0.13)$ &  & $0.34(0.28)$ & $-0.12(0.13)$ & $-0.37(0.39)$ &$-0.07(0.11)$ &1282&28-Feb-23\\
DR Tau & $-4.27(0.04)$ & $-4.54(0.14)$ & $-5.12(0.09)$ &  & $0.80(0.39)$ & $0.46(0.33)$ & $0.41(0.64)$ &$0.58(0.23)$ &1282&4-Mar-23\\
DoAr 25 & $-6.14(0.06)$ & $-5.67(0.12)$ & $-6.17(0.09)$ &  & $-1.72(0.27)$ & $-1.79(0.27)$ & $-1.59(0.68)$ &$-1.74(0.18)$ &1584&16-Aug-23\\
DoAr 33 & $<-6.15$ & $-6.27(0.15)$ & $<-6.80$ &  & $<-1.73$ & $-3.00(0.42)$ & $<-2.79$ &$-3.00(0.42)$ &1584&31-Mar-23\\
Elias 20 & $-5.42(0.06)$ & $-5.44(0.20)$ & $-5.88(0.12)$ &  & $-0.75(0.14)$ & $-1.33(0.41)$ & $-1.03(0.45)$ &$-0.83(0.13)$ &1584&31-Mar-23\\
Elias 24 & $<-4.86$ & $-4.90(0.15)$ & $-5.41(0.06)$ &  & $<-0.00$ & $-0.25(0.32)$ & $-0.14(0.38)$ &$-0.21(0.24)$ &1584&16-Aug-23\\
Elias 27 & $-5.41(0.02)$ & $-5.34(0.08)$ & $-5.86(0.06)$ &  & $-0.73(0.12)$ & $-1.13(0.17)$ & $-1.01(0.39)$ &$-0.87(0.09)$ &1584&16-Mar-24\\
FZ Tau & $-4.68(0.03)$ & $-4.81(0.10)$ & $-5.30(0.07)$ &  & $0.25(0.26)$ & $-0.07(0.23)$ & $0.07(0.46)$ &$0.07(0.16)$ &1549&28-Feb-23\\
GK Tau & $-5.86(0.05)$ & $-5.69(0.11)$ & $-6.00(0.11)$ &  & $-1.34(0.20)$ & $-1.84(0.26)$ & $-1.26(0.53)$ &$-1.50(0.15)$ &1640&28-Feb-23\\
GM Aur & $-5.14(0.01)$ & $-4.99(0.03)$ & $-5.50(0.04)$ &  & $-0.36(0.14)$ & $-0.44(0.10)$ & $-0.32(0.31)$ &$-0.41(0.08)$ &2025&14-Oct-23\\
GO Tau & $-6.54(0.09)$ & $-6.03(0.01)$ & $-6.54(0.05)$ &  & $-2.26(0.40)$ & $-2.52(0.23)$ & $-2.29(1.04)$ &$-2.44(0.20)$ &1640&9-Oct-23\\
GQ Lup & $-5.43(0.06)$ & $-5.17(0.05)$ & $-5.49(0.07)$ &  & $-0.76(0.14)$ & $-0.79(0.12)$ & $-0.29(0.34)$ &$-0.74(0.09)$ &1640&13-Aug-23\\
GW Lup & $<-5.77$ & $-6.08(0.06)$ & $<-6.53$ &  & $<-1.22$ & $-2.62(0.27)$ & $<-2.28$ &$-2.62(0.27)$ &1282&8-Aug-22\\
HD 143006 & $-5.78(0.10)$ & $-5.34(0.02)$ & $-5.63(0.13)$ &  & $-1.23(0.21)$ & $-1.14(0.09)$ & $-0.56(0.37)$ &$-1.13(0.08)$ &1584&4-Apr-23\\
HP Tau & $<-5.53$ & $<-5.62$ & $-5.67(0.15)$ &  & $<-0.90$ & $<-1.70$ & $-0.64(0.40)$ &$-0.64(0.40)$ &1640&27-Feb-23\\
HT Lup & $<-5.17$ & $-5.26(0.10)$ & $-5.62(0.10)$ &  & $<-0.41$ & $-0.96(0.21)$ & $-0.54(0.33)$ &$-0.84(0.18)$ &1584&4-Apr-23\\
IQ Tau & $-5.66(0.03)$ & $-5.50(0.08)$ & $-5.92(0.10)$ &  & $-1.07(0.15)$ & $-1.45(0.19)$ & $-1.11(0.47)$ &$-1.21(0.11)$ &1640&27-Feb-23\\
IRAS-04385 & $<-6.14$ & $-6.01(0.11)$ & $<-6.21$ &  & $<-1.72$ & $-2.48(0.31)$ & $<-1.66$ &$-2.48(0.31)$ &1640&15-Oct-23\\
MY Lup & $<-6.25$ & $-6.17(0.06)$ & $-6.43(0.16)$ &  & $<-1.87$ & $-2.80(0.29)$ & $-2.09(0.97)$ &$-2.74(0.28)$ &1584&13-Aug-23\\
PDS 70 & $<-6.34$ & $<-7.12$ & $-7.09(0.16)$ &  & $<-1.99$ & $<-4.69$ & $-3.34(1.65)$ &$-3.34(1.65)$ &1282&1-Aug-22\\
RU Lup & $-4.61(0.03)$ & $-4.72(0.09)$ & $-5.16(0.09)$ &  & $0.34(0.28)$ & $0.11(0.24)$ & $0.34(0.60)$ &$0.22(0.17)$ &1584&13-Aug-23\\
RY Lup & $<-5.54$ & $-5.28(0.05)$ & $-5.58(0.13)$ &  & $<-0.91$ & $-1.01(0.12)$ & $-0.46(0.37)$ &$-0.96(0.12)$ &1640&13-Aug-23\\
SR 4 & $-4.93(0.02)$ & $-5.06(0.07)$ & $-5.48(0.08)$ &  & $-0.09(0.19)$ & $-0.57(0.16)$ & $-0.28(0.34)$ &$-0.36(0.12)$ &1584&16-Aug-23\\
SY Cha & $<-5.49$ & $-5.82(0.05)$ & $<-6.13$ &  & $<-0.84$ & $-2.09(0.20)$ & $<-1.51$ &$-2.09(0.20)$ &1282&8-Aug-22\\
Sz 98 & $<-5.09$ & $<-5.75$ & $<-6.01$ &  & $<-0.30$ & $<-1.96$ & $<-1.28$ &...&1282&8-Aug-22\\
Sz 114 & $-6.29(0.09)$ & $-6.32(0.21)$ & $<-6.58$ &  & $-1.91(0.33)$ & $-3.09(0.51)$ & $<-2.38$ &$-2.26(0.28)$ &1584&31-Mar-23\\
Sz 129 & $-6.29(0.07)$ & $-5.96(0.06)$ & $-6.42(0.16)$ &  & $-1.92(0.32)$ & $-2.38(0.24)$ & $-2.07(0.96)$ &$-2.21(0.19)$ &1584&31-Mar-23\\
TW Cha & $-5.71(0.06)$ & $-5.67(0.14)$ & $-5.82(0.10)$ &  & $-1.14(0.17)$ & $-1.80(0.32)$ & $-0.93(0.39)$ &$-1.24(0.14)$ &1549&24-Jul-23\\
TW Hya & $-6.09(0.03)$ & $-5.73(0.02)$ & $-6.26(0.06)$ &  & $-1.65(0.25)$ & $-1.92(0.16)$ & $-1.76(0.76)$ &$-1.84(0.13)$ &1282&24-Jan-23\\
V1094 Sco & $<-5.96$ & $-6.00(0.06)$ & $-6.49(0.09)$ &  & $<-1.47$ & $-2.45(0.25)$ & $-2.20(1.00)$ &$-2.43(0.24)$ &1282&9-Aug-22\\
VZ Cha & $-5.38(0.05)$ & $-5.33(0.13)$ & $-5.56(0.08)$ &  & $-0.69(0.13)$ & $-1.12(0.27)$ & $-0.42(0.32)$ &$-0.73(0.11)$ &1549&24-Jul-23\\
WSB 52 & $-5.40(0.05)$ & $-5.19(0.10)$ & $-5.84(0.09)$ &  & $-0.72(0.13)$ & $-0.84(0.22)$ & $-0.97(0.40)$ &$-0.77(0.11)$ &1584&28-Aug-23\\
\enddata
\tablecomments{The \hi\ (7-6) line luminosity has been corrected for water contamination as described in Section \ref{acc_relations}.
The data used here can be found in MAST: \dataset[10.17909/jaye-2n93]{http://dx.doi.org/10.17909/jaye-2n93}. 
JWST Program Information: \\
               ID 1282: PI T.\ Henning, {\it MIRI EC Protoplanetary  and Debris Disks Survey} \\
               ID 1549: PI K.\ Pontoppidan, {\it The deepest search for rare molecules and isotopologues in planet-forming disks} \\
               ID 1584: PI C.\ Salyk, {\it A DSHARP-MIRI Treasury survey of Chemistry in Planet-forming Regions} \\
               ID 1640: PI A.\ Banzatti, {\it The infrared water spectrum as a tracer of pebble delivery to rocky planets} \\
               ID 2025: PI K. Oberg, {\it The Chemistry of Planet Formation: A JWST-ALMA Survey of 4 Planet-Forming Disks}}
\end{deluxetable*}

\section{Summary}

Using near-simultaneous monitoring from three telescopes from the ground and space that covered three strong accretion bursts in the DQ~Tau binary system, we have calibrated the luminosity of three \hi\ lines observed with MIRI-MRS as diagnostics of the stellar accretion luminosity. By measuring the \hi\ line luminosity and accretion luminosity in a single system with predictable accretion bursts, this work significantly reduced the large scatter previously observed in accretion relations from larger samples.  

A fundamental improvement provided in this work, as compared to previous work with IRS, is the calibration of two \hi\ lines that are free from water contamination, avoiding reliance on uncertain model-dependent decompositions. The most reliable line we identify is \hi\ (10-7). The higher resolving power of MIRI also resolves the MIR \hi\ lines for the first time, where large widths support an origin in the accretion process. We also developed a new correction for water contamination for the (7-6) line, previously introduced by \cite{rigliaco2015}, and provide an updated relation for MIRI spectra that clearly resolves adjacent water and \hi\ lines that were blended with the IRS. Finally, we have applied this relation to disks observed in the JDISCS and MINDS surveys, providing an accretion luminosity for these observations. 

The accretion relations provided in this work will allow the community to measure \lacc\ directly from MIRI-MRS spectra, providing a fundamental ingredient for inner disk chemistry and the interpretation of molecular spectra. These direct accretion measurements will also be essential for embedded sources where optical diagnostics are not feasible.

\facilities{\jwst, VLT, Las Cumbres Observatory}
\software{{\tt astropy} \citep{astropy1,astropy2}, 
{\tt photutils} \citep{photutils},
{\tt Astrometry.net} \citep{Langetal2010},
{\tt barycorrpy} \citep{KanodiaAndWright2018},
{\tt scipy} \citep{scipy,scipy2},
{\tt linmix} \citep{Kelly2007},
{\tt LMFIT} \citep{lmfit}, 
{\tt iSLAT} \citep{iSLAT_code},
GitHub Copilot \citep{githubcopilot},
}

\section*{Acknowledgments}

The authors are grateful to the whole team that supports \jwst\ operations and in particular to B. Perriello, B. Sargent, and T. Roman for their help during this \jwst\ monitoring program to maximize the quality of the data and re-schedule two epochs missed by the observatory due to technical issues.

BMT acknowledges support from the Heising-Simons Foundation's 51 Pegasi b Postdoctoral Fellowship in Planetary Astronomy. 

BMT, AB, and JH are funded in part by STScI grant number JWST-GO- 04727.008-A.

CFM, JCW, and HA are funded by the European Union (ERC, WANDA, 101039452). Views and opinions expressed are however those of the author(s) only and do not necessarily reflect those of the European Union or the European Research Council Executive Agency. Neither the European Union nor the granting authority can be held responsible for them.

CHR acknowledges the support of the Deutsche Forschungsgemeinschaft (DFG, German Research Foundation) Research Unit ``Transition discs'' - 325594231.
CHR is grateful for support from the Max Planck Society.

BN acknowledge support from the Large Grant INAF 2022 “YSOs Outflows, Disks and Accretion (YODA): towards a global framework for the evolution of planet forming systems" and from PRIN-MUR 2022 20228JPA3A “The path to star and planet formation in the JWST era (PATH)”

MB has received funding from the European Research Council (ERC) under the European Union’s Horizon 2020 research and innovation programme (PROTOPLANETS, grant agreement No. 101002188).

EF has been partially supported by project
AYA2018-RTI-096188-B-I00 from the Spanish Agencia Estatal de Investigacion and by Grant Agreement 101004719 of the EU projectORP.

A portion of this research was carried out at the Jet Propulsion Laboratory, California Institute of Technology, under a contract with the National Aeronautics and Space Administration (80NM0018D0004).

Some of the data presented in this paper were obtained from the Mikulski Archive for Space Telescopes (MAST) at the Space Telescope Science Institute. 
The specific observations analyzed can be accessed via \dataset[http://dx.doi.org/10.17909/2sqp-pd37]{http://dx.doi.org/10.17909/2sqp-pd37} and \dataset[http://dx.doi.org/10.17909/jaye-2n93]{http://dx.doi.org/10.17909/jaye-2n93}.
STScI is operated by the Association of Universities for Research in Astronomy, Inc., under NASA contract NAS5–26555. Support to MAST for these data is provided by the NASA Office of Space Science via grant NAG5–7584 and by other grants and contracts.

This work makes use of observations from the Las Cumbres Observatory global telescope network.

\appendix

\section{\hi\ lines in absorption} \label{app: HI_abs}
\begin{figure*}
\centering
\includegraphics[width=1\textwidth]{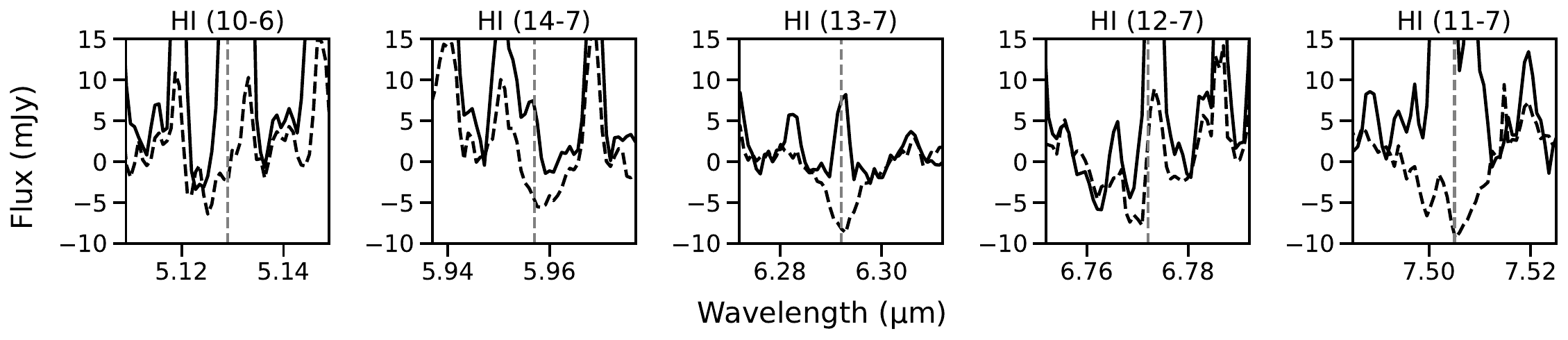}
\caption{Short wavelength \hi\ lines that present absorption during low accretion epochs. Two DQ~Tau epochs are shown in each panel to illustrate a high accretion phase in emission (September 11th, in solid black) and a low accretion phase in absorption (October 7th, dashed black). The same epochs are shown in Figure \ref{fig: HI_series}. \hi\ lines in absorption have also been observed in the low-accretion object MY~Lup in \cite{salyk25}.}
\label{fig: HI_series_abs}
\end{figure*}

Figure \ref{fig: HI_series_abs} shows \hi\ lines observed in this work that turn into broad absorption at low-accretion epochs, and are therefore excluded as accretion tracers. 

\section{Correction of the \hi\ (7-6) line from water contamination} \label{app: correction}
\begin{figure}
\centering
\includegraphics[width=0.45\textwidth]{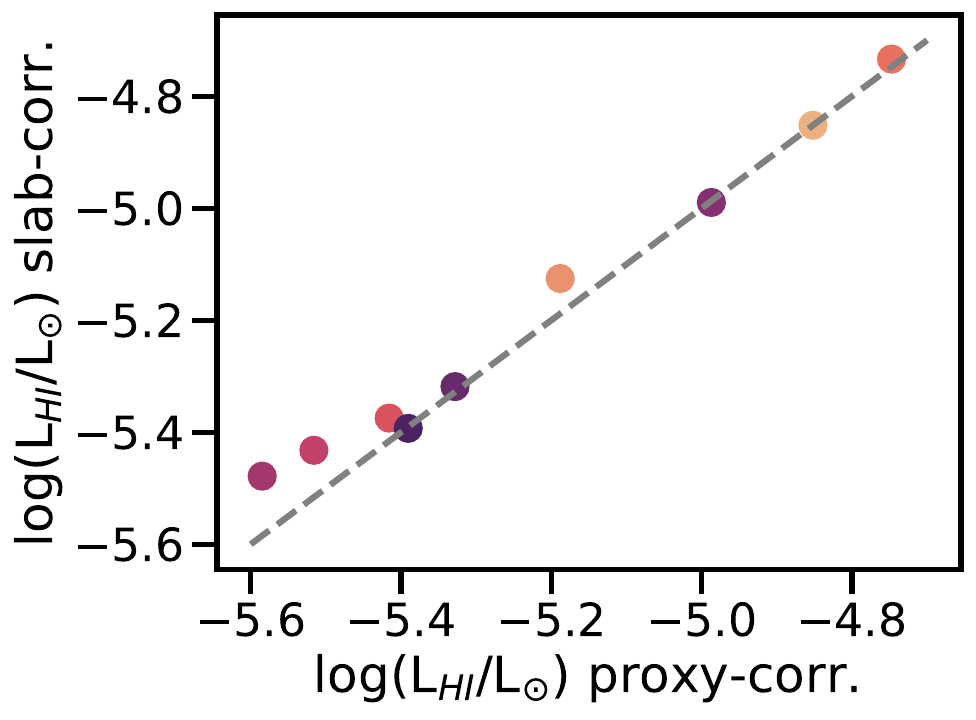}
\caption{Comparison between \hi\ (7-6) line luminosity as corrected from water contamination using the line proxy method vs using a slab model fit (see Appendix \ref{app: correction} for details). The 1:1 relation is shown with a dashed line. }
\label{fig: H2O_correction_comparison}
\end{figure}

The higher resolving power of MIRI spectra supports the development of an improved technique to correct the \hi\ (7-6) line from water contamination, compared to what could be done with the IRS \citep{rigliaco2015}. From the list of isolated water transitions presented in \cite{banzatti24}, we select two nearby transitions of similar upper level energy and strength to the water transition that is blended with \hi\ (7-6) (i.e., $\lambda_{H_2O} = 12.376~\mu$m with $E_u = 4948$~K)\footnote{Note that this is the true transition contaminating \hi\, and not the nearby transition at 12.396~$\mu$m used in \cite{rigliaco2015}, see Figure \ref{fig: HI_series}.}. These transitions are at $\lambda =  11.648~\mu$m with $E_u = 5483$~K and $\lambda =  14.428~\mu$m with $E_u = 4668$~K. We take the average flux of these transitions as a proxy for the flux of the 12.376~$\mu$m and subtract it from the blended line measured between 12.3571 and 12.3815~$\mu$m to obtain the \hi\ (7-6) line flux. We test this correction by fitting a slab model to the hot water component in all epochs of DQ~Tau, and subtracting the best-fit model from the spectra to re-measure the \hi\ (7-6) line flux. To fit the hot component, we use the ``single slab fit" function in iSLAT and fit only the un-blended lines between 10--18~$\mu$m with $E_u > 4500$~K from the list in \cite{banzatti24}. The proxy-corrected \hi\ (7-6) line is $\lesssim 20\%$ off the slab-corrected \hi\ line in all epochs (Figure \ref{fig: H2O_correction_comparison}), whose best-fit models are found at temperatures between 550 and 925~K (these models will be presented in Hyden et al., in prep.). Since disk spectra are typically dominated by a hot (600--900~K) water component at $< 17$~$\mu$m \citep[][Arulanantham et al.\ submitted]{banzatti24,romeromirza24b}, the correction by proxy lines developed in this work should be generally valid for accreting Class II disks. In disks where hot emission is not observed \citep[e.g. MY~Lup,][]{salyk25}, this correction does not change the observed \hi\ (7-6) line flux since the two proxy water lines are not detected (their upper level energy is high enough that they are not populated in absence of hot water).

\begin{thebibliography}{}
\expandafter\ifx\csname natexlab\endcsname\relax\def\natexlab#1{#1}\fi
\providecommand{\url}[1]{\href{#1}{#1}}
\providecommand{\dodoi}[1]{doi:~\href{http://doi.org/#1}{\nolinkurl{#1}}}
\providecommand{\doeprint}[1]{\href{http://ascl.net/#1}{\nolinkurl{http://ascl.net/#1}}}
\providecommand{\doarXiv}[1]{\href{https://arxiv.org/abs/#1}{\nolinkurl{https://arxiv.org/abs/#1}}}

\bibitem[{{Alcal{\'a}} {et~al.}(2014){Alcal{\'a}}, {Natta}, {Manara}, {Spezzi}, {Stelzer}, {Frasca}, {Biazzo}, {Covino}, {Randich}, {Rigliaco}, {Testi}, {Comer{\'o}n}, {Cupani}, \& {D'Elia}}]{Alcalaetal2014}
{Alcal{\'a}}, J.~M., {Natta}, A., {Manara}, C.~F., {et~al.} 2014, \aap, 561, A2, \dodoi{10.1051/0004-6361/201322254}

\bibitem[{{Alcal{\'a}} {et~al.}(2017){Alcal{\'a}}, {Manara}, {Natta}, {Frasca}, {Testi}, {Nisini}, {Stelzer}, {Williams}, {Antoniucci}, {Biazzo}, {Covino}, {Esposito}, {Getman}, \& {Rigliaco}}]{Alcalaetal2017}
{Alcal{\'a}}, J.~M., {Manara}, C.~F., {Natta}, A., {et~al.} 2017, \aap, 600, A20, \dodoi{10.1051/0004-6361/201629929}

\bibitem[{{Ansdell} {et~al.}(2016){Ansdell}, {Gaidos}, {Rappaport}, {Jacobs}, {LaCourse}, {Jek}, {Mann}, {Wyatt}, {Kennedy}, {Williams}, \& {Boyajian}}]{Ansdelletal2016a}
{Ansdell}, M., {Gaidos}, E., {Rappaport}, S.~A., {et~al.} 2016, \apj, 816, 69, \dodoi{10.3847/0004-637X/816/2/69}

\bibitem[{{Argyriou} {et~al.}(2023){Argyriou}, {Glasse}, {Law}, {Labiano}, {{\'A}lvarez-M{\'a}rquez}, {Patapis}, {Kavanagh}, {Gasman}, {Mueller}, {Larson}, {Vandenbussche}, {Glauser}, {Royer}, {Dicken}, {Harkett}, {Sargent}, {Engesser}, {Jones}, {Kendrew}, {Noriega-Crespo}, {Brandl}, {Rieke}, {Wright}, {Lee}, \& {Wells}}]{Argyriou23}
{Argyriou}, I., {Glasse}, A., {Law}, D.~R., {et~al.} 2023, \aap, 675, A111, \dodoi{10.1051/0004-6361/202346489}

\bibitem[{{Artymowicz} \& {Lubow}(1996)}]{Artymowicz&Lubow1996}
{Artymowicz}, P., \& {Lubow}, S.~H. 1996, \apjl, 467, L77, \dodoi{10.1086/310200}

\bibitem[{{Arulanantham} {et~al.}(2024){Arulanantham}, {McClure}, {Pontoppidan}, {Beck}, {Sturm}, {Harsono}, {Boogert}, {Cordiner}, {Dartois}, {Drozdovskaya}, {Espaillat}, {Melnick}, {Noble}, {Palumbo}, {Pendleton}, {Terada}, \& {van Dishoeck}}]{Arulanantham2024}
{Arulanantham}, N., {McClure}, M.~K., {Pontoppidan}, K., {et~al.} 2024, \apjl, 965, L13, \dodoi{10.3847/2041-8213/ad35c9}

\bibitem[{{Aspin} {et~al.}(2010){Aspin}, {Reipurth}, {Herczeg}, \& {Capak}}]{Aspinetal2010}
{Aspin}, C., {Reipurth}, B., {Herczeg}, G.~J., \& {Capak}, P. 2010, \apjl, 719, L50, \dodoi{10.1088/2041-8205/719/1/L50}

\bibitem[{{Astropy Collaboration} {et~al.}(2013){Astropy Collaboration}, {Robitaille}, {Tollerud}, {Greenfield}, {Droettboom}, {Bray}, {Aldcroft}, {Davis}, {Ginsburg}, {Price-Whelan}, {Kerzendorf}, {Conley}, {Crighton}, {Barbary}, {Muna}, {Ferguson}, {Grollier}, {Parikh}, {Nair}, {Unther}, {Deil}, {Woillez}, {Conseil}, {Kramer}, {Turner}, {Singer}, {Fox}, {Weaver}, {Zabalza}, {Edwards}, {Azalee Bostroem}, {Burke}, {Casey}, {Crawford}, {Dencheva}, {Ely}, {Jenness}, {Labrie}, {Lim}, {Pierfederici}, {Pontzen}, {Ptak}, {Refsdal}, {Servillat}, \& {Streicher}}]{astropy2}
{Astropy Collaboration}, {Robitaille}, T.~P., {Tollerud}, E.~J., {et~al.} 2013, \aap, 558, A33, \dodoi{10.1051/0004-6361/201322068}

\bibitem[{{Banzatti} {et~al.}(2014){Banzatti}, {Meyer}, {Manara}, {Pontoppidan}, \& {Testi}}]{banzatti14}
{Banzatti}, A., {Meyer}, M.~R., {Manara}, C.~F., {Pontoppidan}, K.~M., \& {Testi}, L. 2014, \apj, 780, 26, \dodoi{10.1088/0004-637X/780/1/26}

\bibitem[{{Banzatti} {et~al.}(2015){Banzatti}, {Pontoppidan}, {Bruderer}, {Muzerolle}, \& {Meyer}}]{banzatti15}
{Banzatti}, A., {Pontoppidan}, K.~M., {Bruderer}, S., {Muzerolle}, J., \& {Meyer}, M.~R. 2015, \apjl, 798, L16, \dodoi{10.1088/2041-8205/798/1/L16}

\bibitem[{{Banzatti} {et~al.}(2012){Banzatti}, {Meyer}, {Bruderer}, {Geers}, {Pascucci}, {Lahuis}, {Juh{\'a}sz}, {Henning}, \& {{\'A}brah{\'a}m}}]{Banzattietal2012}
{Banzatti}, A., {Meyer}, M.~R., {Bruderer}, S., {et~al.} 2012, \apj, 745, 90, \dodoi{10.1088/0004-637X/745/1/90}

\bibitem[{{Banzatti} {et~al.}(2023{\natexlab{a}}){Banzatti}, {Pontoppidan}, {P{\'e}re Ch{\'a}vez}, {Salyk}, {Diehl}, {Bruderer}, {Herczeg}, {Carmona}, {Pascucci}, {Brittain}, {Jensen}, {Grant}, {van Dishoeck}, {Kamp}, {Bosman}, {{\"O}berg}, {Blake}, {Meyer}, {Gaidos}, {Boogert}, {Rayner}, \& {Wheeler}}]{banzatti23}
{Banzatti}, A., {Pontoppidan}, K.~M., {P{\'e}re Ch{\'a}vez}, J., {et~al.} 2023{\natexlab{a}}, \aj, 165, 72, \dodoi{10.3847/1538-3881/aca80b}

\bibitem[{{Banzatti} {et~al.}(2023{\natexlab{b}}){Banzatti}, {Pontoppidan}, {Carr}, {Jellison}, {Pascucci}, {Najita}, {Romero-Mirza}, {{\"O}berg}, {Kalyaan}, {Pinilla}, {Krijt}, {Long}, {Lambrechts}, {Rosotti}, {Herczeg}, {Salyk}, {Zhang}, {Bergin}, {Ballering}, {Meyer}, {Bruderer}, \& {Jdiscs Collaboration}}]{banzatti23b}
{Banzatti}, A., {Pontoppidan}, K.~M., {Carr}, J.~S., {et~al.} 2023{\natexlab{b}}, \apjl, 957, L22, \dodoi{10.3847/2041-8213/acf5ec}

\bibitem[{{Banzatti} {et~al.}(2025){Banzatti}, {Salyk}, {Pontoppidan}, {Carr}, {Zhang}, {Arulanantham}, {Krijt}, {{\"O}berg}, {Cleeves}, {Najita}, {Pascucci}, {Blake}, {Romero-Mirza}, {Bergin}, {Cieza}, {Pinilla}, {Long}, {Mallaney}, {Xie}, {Waggoner}, {Kaeufer}, \& {Jdiscs Collaboration}}]{banzatti24}
{Banzatti}, A., {Salyk}, C., {Pontoppidan}, K.~M., {et~al.} 2025, \aj, 169, 165, \dodoi{10.3847/1538-3881/ada962}

\bibitem[{{Baraffe} {et~al.}(2015){Baraffe}, {Homeier}, {Allard}, \& {Chabrier}}]{Baraffeetal2015}
{Baraffe}, I., {Homeier}, D., {Allard}, F., \& {Chabrier}, G. 2015, \aap, 577, A42, \dodoi{10.1051/0004-6361/201425481}

\bibitem[{{Basri} {et~al.}(1997){Basri}, {Johns-Krull}, \& {Mathieu}}]{Basrietal1997}
{Basri}, G., {Johns-Krull}, C.~M., \& {Mathieu}, R.~D. 1997, \aj, 114, 781, \dodoi{10.1086/118510}

\bibitem[{{Beuther} {et~al.}(2023){Beuther}, {van Dishoeck}, {Tychoniec}, {Gieser}, {Kavanagh}, {Perotti}, {van Gelder}, {Klaassen}, {Caratti o Garatti}, {Francis}, {Rocha}, {Slavicinska}, {Ray}, {Justtanont}, {Linnartz}, {Waelkens}, {Colina}, {Greve}, {G{\"u}del}, {Henning}, {Lagage}, {Vandenbussche}, {{\"O}stlin}, \& {Wright}}]{Beutheretal2023}
{Beuther}, H., {van Dishoeck}, E.~F., {Tychoniec}, L., {et~al.} 2023, \aap, 673, A121, \dodoi{10.1051/0004-6361/202346167}

\bibitem[{{Bouvier} {et~al.}(2007){Bouvier}, {Alencar}, {Harries}, {Johns-Krull}, \& {Romanova}}]{Bouvieretal2007}
{Bouvier}, J., {Alencar}, S.~H.~P., {Harries}, T.~J., {Johns-Krull}, C.~M., \& {Romanova}, M.~M. 2007, Protostars and Planets V, 479

\bibitem[{Bradley {et~al.}(2024)Bradley, Sip{\H o}cz, Robitaille, Tollerud, Vin{\'{\i}}cius, Deil, Barbary, Wilson, Busko, Donath, G{\"u}nther, Cara, Lim, Me{\ss}linger, Burnett, Conseil, Droettboom, Bostroem, Bray, Bratholm, Jamieson, Ginsburg, Barentsen, Craig, Pascual, Rathi, Perrin, Morris, \& Perren}]{photutils}
Bradley, L., Sip{\H o}cz, B., Robitaille, T., {et~al.} 2024, astropy/photutils: 1.13.0, 1.13.0,  Zenodo, \dodoi{10.5281/zenodo.12585239}

\bibitem[{{Brown} {et~al.}(2013){Brown}, {Baliber}, {Bianco}, {Bowman}, {Burleson}, {Conway}, {Crellin}, {Depagne}, {De Vera}, {Dilday}, {Dragomir}, {Dubberley}, {Eastman}, {Elphick}, {Falarski}, {Foale}, {Ford}, {Fulton}, {Garza}, {Gomez}, {Graham}, {Greene}, {Haldeman}, {Hawkins}, {Haworth}, {Haynes}, {Hidas}, {Hjelstrom}, {Howell}, {Hygelund}, {Lister}, {Lobdill}, {Martinez}, {Mullins}, {Norbury}, {Parrent}, {Paulson}, {Petry}, {Pickles}, {Posner}, {Rosing}, {Ross}, {Sand}, {Saunders}, {Shobbrook}, {Shporer}, {Street}, {Thomas}, {Tsapras}, {Tufts}, {Valenti}, {Vander Horst}, {Walker}, {White}, \& {Willis}}]{Brownetal2013}
{Brown}, T.~M., {Baliber}, N., {Bianco}, F.~B., {et~al.} 2013, \pasp, 125, 1031, \dodoi{10.1086/673168}

\bibitem[{Bushouse {et~al.}(2024)Bushouse, Eisenhamer, Dencheva, Davies, Greenfield, Morrison, Hodge, Simon, Grumm, Droettboom, Slavich, Sosey, Pauly, Miller, Jedrzejewski, Hack, Davis, Crawford, Law, Gordon, Regan, Cara, MacDonald, Bradley, Shanahan, Jamieson, Teodoro, Williams, \& Pena-Guerrero}]{MIRI_pip}
Bushouse, H., Eisenhamer, J., Dencheva, N., {et~al.} 2024, JWST Calibration Pipeline, 1.15.1,  Zenodo, \dodoi{10.5281/zenodo.12692459}

\bibitem[{{Calvet} \& {Gullbring}(1998)}]{Calvet&Gullbring1998}
{Calvet}, N., \& {Gullbring}, E. 1998, \apj, 509, 802, \dodoi{10.1086/306527}

\bibitem[{{Cardelli} {et~al.}(1989){Cardelli}, {Clayton}, \& {Mathis}}]{Cardellietal1989}
{Cardelli}, J.~A., {Clayton}, G.~C., \& {Mathis}, J.~S. 1989, \apj, 345, 245, \dodoi{10.1086/167900}

\bibitem[{{Chen} {et~al.}(2021){Chen}, {Tworek}, {Jun}, {Yuan}, {Ponde de Oliveira Pinto}, {Kaplan}, {Edwards}, {Burda}, {Joseph}, {Brockman}, {Ray}, {Puri}, {Krueger}, {Petrov}, {Khlaaf}, {Sastry}, {Mishkin}, {Chan}, {Gray}, {Ryder}, {Pavlov}, {Power}, {Kaiser}, {Bavarian}, {Winter}, {Tillet}, {Petroski Such}, {Cummings}, {Plappert}, {Chantzis}, {Barnes}, {Herbert-Voss}, {Hebgen Guss}, {Nichol}, {Paino}, {Tezak}, {Tang}, {Babuschkin}, {Balaji}, {Jain}, {Saunders}, {Hesse}, {Carr}, {Leike}, {Achiam}, {Misra}, {Morikawa}, {Radford}, {Knight}, {Brundage}, {Murati}, {Mayer}, {Welinder}, {McGrew}, {Amodei}, {McCandlish}, {Sutskever}, \& {Zaremba}}]{githubcopilot}
{Chen}, M., {Tworek}, J., {Jun}, H., {et~al.} 2021, arXiv e-prints, arXiv:2107.03374, \dodoi{10.48550/arXiv.2107.03374}

\bibitem[{{Claes} {et~al.}(2024){Claes}, {Campbell-White}, {Manara}, {Frasca}, {Natta}, {Alcal{\'a}}, {Armeni}, {Fang}, {Lovell}, {Stelzer}, {Venuti}, {Wyatt}, \& {Queitsch}}]{Claesetal2024}
{Claes}, R.~A.~B., {Campbell-White}, J., {Manara}, C.~F., {et~al.} 2024, \aap, 690, A122, \dodoi{10.1051/0004-6361/202450885}

\bibitem[{{Cody} {et~al.}(2014){Cody}, {Stauffer}, {Baglin}, {Micela}, {Rebull}, {Flaccomio}, {Morales-Calder{\'o}n}, {Aigrain}, {Bouvier}, {Hillenbrand}, {Gutermuth}, {Song}, {Turner}, {Alencar}, {Zwintz}, {Plavchan}, {Carpenter}, {Findeisen}, {Carey}, {Terebey}, {Hartmann}, {Calvet}, {Teixeira}, {Vrba}, {Wolk}, {Covey}, {Poppenhaeger}, {G{\"u}nther}, {Forbrich}, {Whitney}, {Affer}, {Herbst}, {Hora}, {Barrado}, {Holtzman}, {Marchis}, {Wood}, {Medeiros Guimar{\~a}es}, {Lillo Box}, {Gillen}, {McQuillan}, {Espaillat}, {Allen}, {D'Alessio}, \& {Favata}}]{Codyetal2014}
{Cody}, A.~M., {Stauffer}, J., {Baglin}, A., {et~al.} 2014, \aj, 147, 82, \dodoi{10.1088/0004-6256/147/4/82}

\bibitem[{{Colmenares} {et~al.}(2024){Colmenares}, {Bergin}, {Salyk}, {Pontoppidan}, {Arulanantham}, {Calahan}, {Banzatti}, {Andrews}, {Blake}, {Ciesla}, {Green}, {Long}, {Lambrechts}, {Najita}, {Pascucci}, {Pinilla}, {Krijt}, {Trapman}, \& {Jdiscs Collaboration}}]{Colmenares2024}
{Colmenares}, M.~J., {Bergin}, E.~A., {Salyk}, C., {et~al.} 2024, \apj, 977, 173, \dodoi{10.3847/1538-4357/ad8b4f}

\bibitem[{{Costigan} {et~al.}(2014){Costigan}, {Vink}, {Scholz}, {Ray}, \& {Testi}}]{Costiganetal2014}
{Costigan}, G., {Vink}, J.~S., {Scholz}, A., {Ray}, T., \& {Testi}, L. 2014, \mnras, 440, 3444, \dodoi{10.1093/mnras/stu529}

\bibitem[{Czekala {et~al.}(2016)Czekala, Andrews, Torres, Jensen, Stassun, Wilner, \& Latham}]{Czekalaetal2016}
Czekala, I., Andrews, S.~M., Torres, G., {et~al.} 2016, Astrophys. J., 818, 156, \dodoi{10.3847/0004-637X/818/2/156}

\bibitem[{{Dra{\.z}kowska} {et~al.}(2023){Dra{\.z}kowska}, {Bitsch}, {Lambrechts}, {Mulders}, {Harsono}, {Vazan}, {Liu}, {Ormel}, {Kretke}, \& {Morbidelli}}]{Drazkowska2023}
{Dra{\.z}kowska}, J., {Bitsch}, B., {Lambrechts}, M., {et~al.} 2023, in Astronomical Society of the Pacific Conference Series, Vol. 534, Protostars and Planets VII, ed. S.~{Inutsuka}, Y.~{Aikawa}, T.~{Muto}, K.~{Tomida}, \& M.~{Tamura}, 717, \dodoi{10.48550/arXiv.2203.09759}

\bibitem[{{Espaillat} {et~al.}(2022){Espaillat}, {Herczeg}, {Thanathibodee}, {Pittman}, {Calvet}, {Arulanantham}, {France}, {Serna}, {Hern{\'a}ndez}, {K{\'o}sp{\'a}l}, {Walter}, {Frasca}, {Fischer}, {Johns-Krull}, {Schneider}, {Robinson}, {Edwards}, {{\'A}brah{\'a}m}, {Fang}, {Erkal}, {Manara}, {Alcal{\'a}}, {Alecian}, {Alexander}, {Alonso-Santiago}, {Antoniucci}, {Ardila}, {Banzatti}, {Benisty}, {Bergin}, {Biazzo}, {Brice{\~n}o}, {Campbell-White}, {Cleeves}, {Coffey}, {Eisl{\"o}ffel}, {Facchini}, {Fedele}, {Fiorellino}, {Froebrich}, {Gangi}, {Giannini}, {Grankin}, {G{\"u}nther}, {Guo}, {Hartmann}, {Hillenbrand}, {Hinton}, {Kastner}, {Koen}, {Mauc{\'o}}, {Mendigut{\'\i}a}, {Nisini}, {Panwar}, {Principe}, {Robberto}, {Sicilia-Aguilar}, {Valenti}, {Wendeborn}, {Williams}, {Xu}, \& {Yadav}}]{Espaillatetal2022}
{Espaillat}, C.~C., {Herczeg}, G.~J., {Thanathibodee}, T., {et~al.} 2022, \aj, 163, 114, \dodoi{10.3847/1538-3881/ac479d}

\bibitem[{{Fang} {et~al.}(2009){Fang}, {van Boekel}, {Wang}, {Carmona}, {Sicilia-Aguilar}, \& {Henning}}]{Fangetal2009}
{Fang}, M., {van Boekel}, R., {Wang}, W., {et~al.} 2009, \aap, 504, 461, \dodoi{10.1051/0004-6361/200912468}

\bibitem[{{Fiorellino} {et~al.}(2022){Fiorellino}, {Park}, {K{\'o}sp{\'a}l}, \& {{\'A}brah{\'a}m}}]{fiorellino2022}
{Fiorellino}, E., {Park}, S., {K{\'o}sp{\'a}l}, {\'A}., \& {{\'A}brah{\'a}m}, P. 2022, \apj, 928, 81, \dodoi{10.3847/1538-4357/ac4790}

\bibitem[{{Fiorellino} {et~al.}(2023){Fiorellino}, {Tychoniec}, {Cruz-S{\'a}enz de Miera}, {Antoniucci}, {K{\'o}sp{\'a}l}, {Manara}, {Nisini}, \& {Rosotti}}]{Fiorellino2023}
{Fiorellino}, E., {Tychoniec}, {\L}., {Cruz-S{\'a}enz de Miera}, F., {et~al.} 2023, \apj, 944, 135, \dodoi{10.3847/1538-4357/aca320}

\bibitem[{{Fischer} {et~al.}(2023){Fischer}, {Hillenbrand}, {Herczeg}, {Johnstone}, {Kospal}, \& {Dunham}}]{FischeretalPPVII}
{Fischer}, W.~J., {Hillenbrand}, L.~A., {Herczeg}, G.~J., {et~al.} 2023, in Astronomical Society of the Pacific Conference Series, Vol. 534, Protostars and Planets VII, ed. S.~{Inutsuka}, Y.~{Aikawa}, T.~{Muto}, K.~{Tomida}, \& M.~{Tamura}, 355, \dodoi{10.48550/arXiv.2203.11257}

\bibitem[{{Franceschi} {et~al.}(2024){Franceschi}, {Henning}, {Tabone}, {Perotti}, {Caratti o Garatti}, {Bettoni}, {van Dishoeck}, {Kamp}, {Absil}, {G{\"u}del}, {Olofsson}, {Waters}, {Arabhavi}, {Christiaens}, {Gasman}, {Grant}, {Jang}, {Rodgers-Lee}, {Samland}, {Schwarz}, {Temmink}, {Barrado}, {Boccaletti}, {Geers}, {Lagage}, {Pantin}, {Ray}, {Scheithauer}, {Vandenbussche}, \& {Wright}}]{Franceschietal2024}
{Franceschi}, R., {Henning}, T., {Tabone}, B., {et~al.} 2024, \aap, 687, A96, \dodoi{10.1051/0004-6361/202348034}

\bibitem[{{Freudling} {et~al.}(2013){Freudling}, {Romaniello}, {Bramich}, {Ballester}, {Forchi}, {Garc{\'\i}a-Dabl{\'o}}, {Moehler}, \& {Neeser}}]{reflex}
{Freudling}, W., {Romaniello}, M., {Bramich}, D.~M., {et~al.} 2013, \aap, 559, A96, \dodoi{10.1051/0004-6361/201322494}

\bibitem[{{Gardner} {et~al.}(2023){Gardner}, {Mather}, {Abbott}, {Abell}, {Abernathy}, {Abney}, {Abraham}, {Abraham}, {Abul-Huda}, {Acton}, {Adams}, {Adams}, {Adler}, {Adriaensen}, {Aguilar}, {Ahmed}, {Ahmed}, {Ahmed}, {Albat}, {Albert}, {Alberts}, {Aldridge}, {Allen}, {Allen}, {Altenburg}, {Altunc}, {Alvarez}, {{\'A}lvarez-M{\'a}rquez}, {de Oliveira}, {Ambrose}, {Anandakrishnan}, {Andersen}, {Anderson}, {Anderson}, {Anderson}, {Anderson}, {Aprea}, {Archer}, {Arenberg}, {Argyriou}, {Arribas}, {Artigau}, {Arvai}, {Atcheson}, {Atkinson}, {Averbukh}, {Aymergen}, {Bacinski}, {Baggett}, {Bagnasco}, {Baker}, {Balzano}, {Banks}, {Baran}, {Barker}, {Barrett}, {Barringer}, {Barto}, {Bast}, {Baudoz}, {Baum}, {Beatty}, {Beaulieu}, {Bechtold}, {Beck}, {Beddard}, {Beichman}, {Bellagama}, {Bely}, {Berger}, {Bergeron}, {Bernier}, {Bertch}, {Beskow}, {Betz}, {Biagetti}, {Birkmann}, {Bjorklund}, {Blackwood}, {Blazek}, {Blossfeld}, {Bluth}, {Boccaletti}, {Boegner}, {Bohlin}, {Boia}, {B{\"o}ker}, {Bonaventura}, {Bond},
  {Bosley}, {Boucarut}, {Bouchet}, {Bouwman}, {Bower}, {Bowers}, {Bowers}, {Boyce}, {Boyer}, {Boyer}, {Boyer}, {Boyer}, {Bradley}, {Brady}, {Brandl}, {Brannen}, {Breda}, {Bremmer}, {Brennan}, {Bresnahan}, {Bright}, {Broiles}, {Bromenschenkel}, {Brooks}, {Brooks}, {Brown}, {Brown}, {Brown}, {Bruce}, {Bryson}, {Bujanda}, {Bullock}, {Bunker}, {Bureo}, {Burt}, {Bush}, {Bushouse}, {Bussman}, {Cabaud}, {Cale}, {Calhoon}, {Calvani}, {Canipe}, {Caputo}, {Cara}, {Carey}, {Case}, {Cesari}, {Cetorelli}, {Chance}, {Chandler}, {Chaney}, {Chapman}, {Charlot}, {Chayer}, {Cheezum}, {Chen}, {Chen}, {Cherinka}, {Chichester}, {Chilton}, {Chittiraibalan}, {Clampin}, {Clark}, {Clark}, {Clark}, {Claybrooks}, {Cleveland}, {Cohen}, {Cohen}, {Col{\'o}n}, {Coleman}, {Colina}, {Comber}, {Comeau}, {Comer}, {Reis}, {Connolly}, {Conroy}, {Contos}, {Contreras}, {Cook}, {Cooper}, {Cooper}, {Correia}, {Correnti}, {Cossou}, {Costanza}, {Coulais}, {Cox}, {Coyle}, {Cracraft}, {Crew}, {Curtis}, {Cusveller}, {Maciel}, {Dailey}, {Daugeron},
  {Davidson}, {Davies}, {Davis}, {Davis}, {Day}, {de Chambure}, {de Jong}, {De Marchi}, {Dean}, {Decker}, {Delisa}, {Dell}, {Dellagatta}, {Dembinska}, {Demosthenes}, {Dencheva}, {Deneu}, {DePriest}, {Deschenes}, {Dethienne}, {Detre}, {Diaz}, {Dicken}, {DiFelice}, {Dillman}, {Disharoon}, {Dixon}, {Doggett}, {Dominguez}, {Donaldson}, {Doria-Warner}, {Santos}, {Doty}, {Douglas}, {Doyon}, {Dressler}, {Driggers}, {Driggers}, {Dunn}, {DuPrie}, {Dupuis}, {Durning}, {Dutta}, {Earl}, {Eccleston}, {Ecobichon}, {Egami}, {Ehrenwinkler}, {Eisenhamer}, {Eisenhower}, {Eisenstein}, {El Hamel}, {Elie}, {Elliott}, {Elliott}, {Engesser}, {Espinoza}, {Etienne}, {Etxaluze}, {Evans}, {Fabreguettes}, {Falcolini}, {Falini}, {Fatig}, {Feeney}, {Feinberg}, {Fels}, {Ferdous}, {Ferguson}, {Ferrarese}, {Ferreira}, {Ferruit}, {Ferry}, {Filippazzo}, {Firre}, {Fix}, {Flagey}, {Flanagan}, {Fleming}, {Florian}, {Flynn}, {Foiadelli}, {Fontaine}, {Fontanella}, {Forshay}, {Fortner}, {Fox}, {Framarini}, {Francisco}, {Franck}, {Franx}, {Franz},
  {Friedman}, {Friend}, {Frost}, {Fu}, {Fullerton}, {Gaillard}, {Galkin}, {Gallagher}, {Galyer}, {Garc{\'\i}a Mar{\'\i}n}, {Gardner}, {Garland}, {Garrett}, {Gasman}, {G{\'a}sp{\'a}r}, {Gastaud}, {Gaudreau}, {Gauthier}, {Geers}, {Geithner}, {Gennaro}, {Gerber}, {Gereau}, {Giampaoli}, {Giardino}, {Gibbons}, {Gilbert}, {Gilman}, {Girard}, {Giuliano}, {Gkountis}, {Glasse}, {Glassmire}, {Glauser}, {Glazer}, {Goldberg}, {Golimowski}, {Gonzaga}, {Gordon}, {Gordon}, {Goudfrooij}, {Gough}, {Graham}, {Grau}, {Green}, {Greene}, {Greene}, {Greenfield}, {Greenhouse}, {Greve}, {Greville}, {Grimaldi}, {Groe}, {Groebner}, {Grumm}, {Grundy}, {G{\"u}del}, {Guillard}, {Guldalian}, {Gunn}, {Gurule}, {Gutman}, {Guy}, {Guyot}, {Hack}, {Haderlein}, {Hagan}, {Hagedorn}, {Hainline}, {Haley}, {Hami}, {Hamilton}, {Hammann}, {Hammel}, {Hanley}, {Hansen}, {Hardy}, {Harnisch}, {Harr}, {Harris}, {Hart}, {Hartig}, {Hasan}, {Hashim}, {Hashimoto}, {Haskins}, {Hawkins}, {Hayden}, {Hayden}, {Healy}, {Hecht}, {Heeg}, {Hejal}, {Helm},
  {Hengemihle}, {Henning}, {Henry}, {Henry}, {Henshaw}, {Hernandez}, {Herrington}, {Heske}, {Hesman}, {Hickey}, {Hilbert}, {Hines}, {Hinz}, {Hirsch}, {Hitcho}, {Hodapp}, {Hodge}, {Hoffman}, {Holfeltz}, {Holler}, {Hoppa}, {Horner}, {Howard}, {Howard}, {Huber}, {Hunkeler}, {Hunter}, {Hunter}, {Hurd}, {Hurst}, {Hutchings}, {Hylan}, {Ignat}, {Illingworth}, {Irish}, {Isaacs}, {Jackson}, {Jaffe}, {Jahic}, {Jahromi}, {Jakobsen}, {James}, {James}, {James}, {Jamieson}, {Jandra}, {Jayawardhana}, {Jedrzejewski}, {Jeffers}, {Jensen}, {Joanne}, {Johns}, {Johnson}, {Johnson}, {Johnson}, {Johnson}, {Johnson}, {Johnson}, {Johnstone}, {Jollet}, {Jones}, {Jones}, {Jones}, {Jones}, {Jones}, {Jordan}, {Jordan}, {Jue}, {Jurkowski}, {Justis}, {Justtanont}, {Kaleida}, {Kalirai}, {Kalmanson}, {Kaltenegger}, {Kammerer}, {Kan}, {Kanarek}, {Kao}, {Karakla}, {Karl}, {Kassin}, {Kauffman}, {Kavanagh}, {Kelley}, {Kelly}, {Kendrew}, {Kennedy}, {Kenny}, {Keski-Kuha}, {Keyes}, {Khan}, {Kidwell}, {Kimble}, {King}, {King}, {Kinzel}, {Kirk},
  {Kirkpatrick}, {Klaassen}, {Klingemann}, {Klintworth}, {Knapp}, {Knight}, {Knollenberg}, {Knutsen}, {Koehler}, {Koekemoer}, {Kofler}, {Kontson}, {Kovacs}, {Kozhurina-Platais}, {Krause}, {Kriss}, {Krist}, {Kristoffersen}, {Krogel}, {Krueger}, {Kulp}, {Kumari}, {Kwan}, {Kyprianou}, {Labador}, {Labiano}, {Lafreni{\`e}re}, {Lagage}, {Laidler}, {Laine}, {Laird}, {Lajoie}, {Lallo}, {Lam}, {LaMassa}, {Lambros}, {Lampenfield}, {Lander}, {Langston}, {Larson}, {Larson}, {LaVerghetta}, {Law}, {Lawrence}, {Lee}, {Lee}, {Lee}, {Leisenring}, {Leveille}, {Levenson}, {Levi}, {Levine}, {Lewis}, {Lewis}, {Lewis}, {Libralato}, {Lidon}, {Liebrecht}, {Lightsey}, {Lilly}, {Lim}, {Lim}, {Ling}, {Link}, {Link}, {Lipinski}, {Liu}, {Lo}, {Lobmeyer}, {Logue}, {Long}, {Long}, {Long}, {Long}, {L{\'o}pez-Caniego}, {Lotz}, {Love-Pruitt}, {Lubskiy}, {Luers}, {Luetgens}, {Luevano}, {Lui}, {Lund}, {Lundquist}, {Lunine}, {L{\"u}tzgendorf}, {Lynch}, {MacDonald}, {MacDonald}, {Macias}, {Macklis}, {Maghami}, {Maharaja}, {Maiolino},
  {Makrygiannis}, {Malla}, {Malumuth}, {Manjavacas}, {Marini}, {Marrione}, {Marston}, {Martel}, {Martin}, {Martin}, {Martinez}, {Maschmann}, {Masci}, {Masetti}, {Maszkiewicz}, {Matthews}, {Matuskey}, {McBrayer}, {McCarthy}, {McCaughrean}, {McClare}, {McClare}, {McCloskey}, {McClurg}, {McCoy}, {McElwain}, {McGregor}, {McGuffey}, {McKay}, {McKenzie}, {McLean}, {McMaster}, {McNeil}, {De Meester}, {Mehalick}, {Meixner}, {Mel{\'e}ndez}, {Menzel}, {Menzel}, {Merz}, {Mesterharm}, {Meyer}, {Meyett}, {Meza}, {Midwinter}, {Milam}, {Miller}, {Miller}, {Miskey}, {Misselt}, {Mitchell}, {Mohan}, {Montoya}, {Moran}, {Morishita}, {Moro-Mart{\'\i}n}, {Morrison}, {Morrison}, {Morse}, {Moschos}, {Moseley}, {Mosier}, {Mosner}, {Mountain}, {Muckenthaler}, {Mueller}, {Mueller}, {Muhiem}, {M{\"u}hlmann}, {Mullally}, {Mullen}, {Munger}, {Murphy}, {Murray}, {Muzerolle}, {Mycroft}, {Myers}, {Myers}, {Myers}, {Myers}, {Myrick}, {Nagle}, {Nayak}, {Naylor}, {Neff}, {Nelan}, {Nella}, {Nguyen}, {Nguyen}, {Nickson}, {Nidhiry}, {Niedner},
  {Nieto-Santisteban}, {Nikolov}, {Nishisaka}, {Noriega-Crespo}, {Nota}, {O'Mara}, {Oboryshko}, {O'Brien}, {Ochs}, {Offenberg}, {Ogle}, {Ohl}, {Olmsted}, {Osborne}, {O'Shaughnessy}, {{\"O}stlin}, {O'Sullivan}, {Otor}, {Ottens}, {Ouellette}, {Outlaw}, {Owens}, {Pacifici}, {Page}, {Paranilam}, {Park}, {Parrish}, {Paschal}, {Patapis}, {Patel}, {Patrick}, {Pattishall}, {Paul}, {Paul}, {Pauly}, {Pavlovsky}, {Pe{\~n}a-Guerrero}, {Pedder}, {Peek}, {Pelham}, {Penanen}, {Perriello}, {Perrin}, {Perrine}, {Perrygo}, {Peslier}, {Petach}, {Peterson}, {Pfarr}, {Pierson}, {Pietraszkiewicz}, {Pilchen}, {Pipher}, {Pirzkal}, {Pitman}, {Player}, {Plesha}, {Plitzke}, {Pohner}, {Poletis}, {Pollizzi}, {Polster}, {Pontius}, {Pontoppidan}, {Porges}, {Potter}, {Prescott}, {Proffitt}, {Pueyo}, {Quispe Neira}, {Radich}, {Rager}, {Rameau}, {Ramey}, {Alarcon}, {Rampini}, {Rapp}, {Rashford}, {Rauscher}, {Ravindranath}, {Rawle}, {Rawlings}, {Ray}, {Regan}, {Rehm}, {Rehm}, {Reid}, {Reis}, {Renk}, {Reoch}, {Ressler}, {Rest}, {Reynolds},
  {Richon}, {Richon}, {Ridgaway}, {Riedel}, {Rieke}, {Rieke}, {Rifelli}, {Rigby}, {Riggs}, {Ringel}, {Ritchie}, {Rix}, {Robberto}, {Robinson}, {Robinson}, {Robinson}, {Rock}, {Rodriguez}, {Rodr{\'\i}guez del Pino}, {Roellig}, {Rohrbach}, {Roman}, {Romelfanger}, {Romo}, {Rosales}, {Rose}, {Roteliuk}, {Roth}, {Rothwell}, {Rouzaud}, {Rowe}, {Rowlands}, {Roy}, {Royer}, {Rui}, {Rumler}, {Rumpl}, {Russ}, {Ryan}, {Ryan}, {Saad}, {Sabata}, {Sabatino}, {Sabbi}, {Sabelhaus}, {Sabia}, {Sahu}, {Saif}, {Salvignol}, {Samara-Ratna}, {Samuelson}, {Sanders}, {Sappington}, {Sargent}, {Sauer}, {Savadkin}, {Sawicki}, {Schappell}, {Scheffer}, {Scheithauer}, {Scherer}, {Schiff}, {Schlawin}, {Schmeitzky}, {Schmitz}, {Schmude}, {Schneider}, {Schreiber}, {Schroeven-Deceuninck}, {Schultz}, {Schwab}, {Schwartz}, {Scoccimarro}, {Scott}, {Scott}, {Seaton}, {Seely}, {Seery}, {Seidleck}, {Sembach}, {Shanahan}, {Shaughnessy}, {Shaw}, {Shay}, {Sheehan}, {Sheth}, {Shih}, {Shivaei}, {Siegel}, {Sienkiewicz}, {Simmons}, {Simon}, {Sirianni},
  {Sivaramakrishnan}, {Slade}, {Sloan}, {Slocum}, {Slowinski}, {Smith}, {Smith}, {Smith}, {Smith}, {Smith}, {Smith}, {Smolik}, {Soderblom}, {Sohn}, {Sokol}, {Sonneborn}, {Sontag}, {Sooy}, {Soummer}, {Southwood}, {Spain}, {Sparmo}, {Speer}, {Spencer}, {Sprofera}, {Stallcup}, {Stanley}, {Stansberry}, {Stark}, {Starr}, {Stassi}, {Steck}, {Steeley}, {Stephens}, {Stephenson}, {Stewart}, {Stiavelli}, {}, {Strada}, {Straughn}, {Streetman}, {Strickland}, {Strobele}, {Stuhlinger}, {Stys}, {Such}, {Sukhatme}, {Sullivan}, {Sullivan}, {Sumner}, {Sun}, {Sunnquist}, {Swade}, {Swam}, {Swenton}, {Swoish}, {Tam Litten}, {Tamas}, {Tao}, {Taylor}, {Taylor}, {Plate}, {Van Tea}, {Teague}, {Telfer}, {Temim}, {Texter}, {Thatte}, {Thompson}, {Thompson}, {Thomson}, {Thronson}, {Tierney}, {Tikkanen}, {Tinnin}, {Tippet}, {Todd}, {Tran}, {Trauger}, {Trejo}, {Vinh Truong}, {Tsukamoto}, {Tufail}, {Tumlinson}, {Tustain}, {Tyra}, {Ubeda}, {Underwood}, {Uzzo}, {Vaclavik}, {Valenduc}, {Valenti}, {Van Campen}, {van de Wetering}, {Van Der
  Marel}, {van Haarlem}, {Vandenbussche}, {van Dishoeck}, {Vanterpool}, {Vernoy}, {Vila Costas}, {Volk}, {Voorzaat}, {Voyton}, {Vydra}, {Waddy}, {Waelkens}, {Wahlgren}, {Walker}, {Wander}, {Warfield}, {Warner}, {Wasiak}, {Wasiak}, {Wehner}, {Weiler}, {Weilert}, {Weiss}, {Wells}, {Welty}, {Wheate}, {Wheeler}, {White}, {Whitehouse}, {Whiteleather}, {Whitman}, {Williams}, {Willmer}, {Willott}, {Willoughby}, {Wilson}, {Wilson}, {Wilson}, {Windhorst}, {Wislowski}, {Wolfe}, {Wolfe}, {Wolff}, {Wondel}, {Woo}, {Woods}, {Worden}, {Workman}, {Wright}, {Wu}, {Wu}, {Wun}, {Wymer}, {Yadetie}, {Yan}, {Yang}, {Yates}, {Yeager}, {Yerger}, {Young}, {Young}, {Yu}, {Yu}, {Zak}, {Zeidler}, {Zepp}, {Zhou}, {Zincke}, {Zonak}, \& {Zondag}}]{Gardner23}
{Gardner}, J.~P., {Mather}, J.~C., {Abbott}, R., {et~al.} 2023, \pasp, 135, 068001, \dodoi{10.1088/1538-3873/acd1b5}

\bibitem[{{Gasman} {et~al.}(2023){Gasman}, {van Dishoeck}, {Grant}, {Temmink}, {Tabone}, {Henning}, {Kamp}, {G{\"u}del}, {Lagage}, {Perotti}, {Christiaens}, {Samland}, {Arabhavi}, {Argyriou}, {Abergel}, {Absil}, {Barrado}, {Boccaletti}, {Bouwman}, {Caratti o Garatti}, {Geers}, {Glauser}, {Guadarrama}, {Jang}, {Kanwar}, {Lahuis}, {Morales-Calder{\'o}n}, {Mueller}, {Nehm{\'e}}, {Olofsson}, {Pantin}, {Pawellek}, {Ray}, {Rodgers-Lee}, {Scheithauer}, {Schreiber}, {Schwarz}, {Vandenbussche}, {Vlasblom}, {Waters}, {Wright}, {Colina}, {Greve}, \& {{\"O}stlin}}]{gasman23}
{Gasman}, D., {van Dishoeck}, E.~F., {Grant}, S.~L., {et~al.} 2023, \aap, 679, A117, \dodoi{10.1051/0004-6361/202347005}

\bibitem[{{Gasman} {et~al.}(2025){Gasman}, {Temmink}, {van Dishoeck}, {Kurtovic}, {Grant}, {Sellek}, {Tabone}, {Henning}, {Kamp}, {G{\"u}del}, {Barrado}, {Garatti}, {Glauser}, {Waters}, {Arabhavi}, {Jang}, {Kanwar}, {Lienert}, {Perotti}, {Schwarz}, \& {Vlasblom}}]{gasman25}
{Gasman}, D., {Temmink}, M., {van Dishoeck}, E.~F., {et~al.} 2025, arXiv e-prints, arXiv:2501.04587, \dodoi{10.48550/arXiv.2501.04587}

\bibitem[{{Getman} {et~al.}(2023){Getman}, {K{\'o}sp{\'a}l}, {Arulanantham}, {Semenov}, {Smirnov-Pinchukov}, \& {van Terwisga}}]{Getmanetal2023}
{Getman}, K.~V., {K{\'o}sp{\'a}l}, {\'A}., {Arulanantham}, N., {et~al.} 2023, \apj, 959, 98, \dodoi{10.3847/1538-4357/ad054c}

\bibitem[{{Grant} {et~al.}(2023){Grant}, {van Dishoeck}, {Tabone}, {Gasman}, {Henning}, {Kamp}, {G{\"u}del}, {Lagage}, {Bettoni}, {Perotti}, {Christiaens}, {Samland}, {Arabhavi}, {Argyriou}, {Abergel}, {Absil}, {Barrado}, {Boccaletti}, {Bouwman}, {o Garatti}, {Geers}, {Glauser}, {Guadarrama}, {Jang}, {Kanwar}, {Lahuis}, {Morales-Calder{\'o}n}, {Mueller}, {Nehm{\'e}}, {Olofsson}, {Pantin}, {Pawellek}, {Ray}, {Rodgers-Lee}, {Scheithauer}, {Schreiber}, {Schwarz}, {Temmink}, {Vandenbussche}, {Vlasblom}, {Waters}, {Wright}, {Colina}, {Greve}, {Justannont}, \& {{\"O}stlin}}]{grant23}
{Grant}, S.~L., {van Dishoeck}, E.~F., {Tabone}, B., {et~al.} 2023, \apjl, 947, L6, \dodoi{10.3847/2041-8213/acc44b}

\bibitem[{{Grant} {et~al.}(2024){Grant}, {Kurtovic}, {van Dishoeck}, {Henning}, {Kamp}, {Nowacki}, {Perraut}, {Banzatti}, {Temmink}, {Christiaens}, {Samland}, {Gasman}, {Tabone}, {G{\"u}del}, {Lagage}, {Arabhavi}, {Barrado}, {Garatti}, {Glauser}, {Jang}, {Kanwar}, {Lahuis}, {Morales-Calder{\'o}n}, {Olofsson}, {Perotti}, {Schwarz}, {Vlasblom}, {Garcia Lopez}, \& {Long}}]{grant24}
{Grant}, S.~L., {Kurtovic}, N.~T., {van Dishoeck}, E.~F., {et~al.} 2024, arXiv e-prints, arXiv:2406.10217, \dodoi{10.48550/arXiv.2406.10217}

\bibitem[{{Gullbring} {et~al.}(1998){Gullbring}, {Hartmann}, {Briceno}, \& {Calvet}}]{Gullbringetal1998}
{Gullbring}, E., {Hartmann}, L., {Briceno}, C., \& {Calvet}, N. 1998, \apj, 492, 323, \dodoi{10.1086/305032}

\bibitem[{{Hartmann} {et~al.}(2016){Hartmann}, {Herczeg}, \& {Calvet}}]{Hartmannetal2016}
{Hartmann}, L., {Herczeg}, G., \& {Calvet}, N. 2016, \araa, 54, 135, \dodoi{10.1146/annurev-astro-081915-023347}

\bibitem[{{Hartmann} {et~al.}(1994){Hartmann}, {Hewett}, \& {Calvet}}]{Hartmannetal1994}
{Hartmann}, L., {Hewett}, R., \& {Calvet}, N. 1994, \apj, 426, 669, \dodoi{10.1086/174104}

\bibitem[{{Henning} {et~al.}(2024){Henning}, {Kamp}, {Samland}, {Arabhavi}, {Kanwar}, {van Dishoeck}, {G{\"u}del}, {Lagage}, {Waelkens}, {Abergel}, {Absil}, {Barrado}, {Boccaletti}, {Bouwman}, {Caratti o Garatti}, {Geers}, {Glauser}, {Lahuis}, {Mueller}, {Nehm{\'e}}, {Olofsson}, {Pantin}, {Ray}, {Scheithauer}, {Vandenbussche}, {Waters}, {Wright}, {Argyriou}, {Christiaens}, {Franceschi}, {Gasman}, {Grant}, {Guadarrama}, {Jang}, {Morales-Calder{\'o}n}, {Pawellek}, {Perotti}, {Rodgers-Lee}, {Schreiber}, {Schwarz}, {Tabone}, {Temmink}, {Vlasblom}, {Colina}, {Greve}, \& {{\"O}stlin}}]{MINDS24}
{Henning}, T., {Kamp}, I., {Samland}, M., {et~al.} 2024, \pasp, 136, 054302, \dodoi{10.1088/1538-3873/ad3455}

\bibitem[{{Herczeg} \& {Hillenbrand}(2008)}]{HerczegHillenbrand2008}
{Herczeg}, G.~J., \& {Hillenbrand}, L.~A. 2008, \apj, 681, 594, \dodoi{10.1086/586728}

\bibitem[{{Honeycutt}(1992)}]{Honeycutt1992}
{Honeycutt}, R.~K. 1992, \pasp, 104, 435, \dodoi{10.1086/133015}

\bibitem[{{Ingleby} {et~al.}(2013){Ingleby}, {Calvet}, {Herczeg}, {Blaty}, {Walter}, {Ardila}, {Alexander}, {Edwards}, {Espaillat}, {Gregory}, {Hillenbrand}, \& {Brown}}]{Inglebyetal2013}
{Ingleby}, L., {Calvet}, N., {Herczeg}, G., {et~al.} 2013, \apj, 767, 112, \dodoi{10.1088/0004-637X/767/2/112}

\bibitem[{{Jellison} {et~al.}(2024){Jellison}, {Johnson}, {Banzatti}, \& {Bruderer}}]{iSLAT}
{Jellison}, E., {Johnson}, M., {Banzatti}, A., \& {Bruderer}, S. 2024, arXiv e-prints, arXiv:2402.04060, \dodoi{10.48550/arXiv.2402.04060}

\bibitem[{Johnson {et~al.}(2024)Johnson, Banzatti, Fuller, \& Jellison}]{iSLAT_code}
Johnson, M., Banzatti, A., Fuller, J., \& Jellison, E. 2024, spexod/iSLAT: Second release, v4.03,  Zenodo, \dodoi{10.5281/zenodo.12167853}

\bibitem[{Jones {et~al.}(2001)Jones, Oliphant, Peterson, {et~al.}}]{scipy}
Jones, E., Oliphant, T., Peterson, P., {et~al.} 2001, {SciPy}: Open source scientific tools for {Python}.
\newblock \url{http://www.scipy.org/}

\bibitem[{{Kanodia} \& {Wright}(2018)}]{KanodiaAndWright2018}
{Kanodia}, S., \& {Wright}, J.~T. 2018, {Barycorrpy: Barycentric velocity calculation and leap second management}.
\newblock \doeprint{1808.001}

\bibitem[{{Kanwar} {et~al.}(2024){Kanwar}, {Kamp}, {Jang}, {Waters}, {van Dishoeck}, {Christiaens}, {Arabhavi}, {Henning}, {G{\"u}del}, {Woitke}, {Absil}, {Barrado}, {Caratti o Garatti}, {Glauser}, {Lahuis}, {Scheithauer}, {Vandenbussche}, {Gasman}, {Grant}, {Kurtovic}, {Perotti}, {Tabone}, \& {Temmink}}]{Kanwar2024}
{Kanwar}, J., {Kamp}, I., {Jang}, H., {et~al.} 2024, \aap, 689, A231, \dodoi{10.1051/0004-6361/202450078}

\bibitem[{{Kelly}(2007)}]{Kelly2007}
{Kelly}, B.~C. 2007, \apj, 665, 1489, \dodoi{10.1086/519947}

\bibitem[{{K{\'o}sp{\'a}l} {et~al.}(2023){K{\'o}sp{\'a}l}, {{\'A}brah{\'a}m}, {Diehl}, {Banzatti}, {Bouwman}, {Chen}, {Cruz-S{\'a}enz de Miera}, {Green}, {Henning}, \& {Rab}}]{Kospaletal2023}
{K{\'o}sp{\'a}l}, {\'A}., {{\'A}brah{\'a}m}, P., {Diehl}, L., {et~al.} 2023, \apjl, 945, L7, \dodoi{10.3847/2041-8213/acb58a}

\bibitem[{{Kwan} \& {Fischer}(2011)}]{KwanFischer}
{Kwan}, J., \& {Fischer}, W. 2011, \mnras, 411, 2383, \dodoi{10.1111/j.1365-2966.2010.17863.x}

\bibitem[{{Lang} {et~al.}(2010){Lang}, {Hogg}, {Mierle}, {Blanton}, \& {Roweis}}]{Langetal2010}
{Lang}, D., {Hogg}, D.~W., {Mierle}, K., {Blanton}, M., \& {Roweis}, S. 2010, \aj, 139, 1782, \dodoi{10.1088/0004-6256/139/5/1782}

\bibitem[{{Le Gouellec} {et~al.}(2024){Le Gouellec}, {Greene}, {Hillenbrand}, \& {Yates}}]{LeGouellec2024}
{Le Gouellec}, V. J.~M., {Greene}, T.~P., {Hillenbrand}, L.~A., \& {Yates}, Z. 2024, \apj, 966, 91, \dodoi{10.3847/1538-4357/ad2935}

\bibitem[{{Manara} {et~al.}(2023){Manara}, {Ansdell}, {Rosotti}, {Hughes}, {Armitage}, {Lodato}, \& {Williams}}]{ManaraetalPPVII}
{Manara}, C.~F., {Ansdell}, M., {Rosotti}, G.~P., {et~al.} 2023, in Astronomical Society of the Pacific Conference Series, Vol. 534, Protostars and Planets VII, ed. S.~{Inutsuka}, Y.~{Aikawa}, T.~{Muto}, K.~{Tomida}, \& M.~{Tamura}, 539, \dodoi{10.48550/arXiv.2203.09930}

\bibitem[{{Manara} {et~al.}(2013{\natexlab{a}}){Manara}, {Beccari}, {Da Rio}, {De Marchi}, {Natta}, {Ricci}, {Robberto}, \& {Testi}}]{Manaraetal2013fitter}
{Manara}, C.~F., {Beccari}, G., {Da Rio}, N., {et~al.} 2013{\natexlab{a}}, \aap, 558, A114, \dodoi{10.1051/0004-6361/201321866}

\bibitem[{{Manara} {et~al.}(2017){Manara}, {Frasca}, {Alcal{\'a}}, {Natta}, {Stelzer}, \& {Testi}}]{Manaraetal2017}
{Manara}, C.~F., {Frasca}, A., {Alcal{\'a}}, J.~M., {et~al.} 2017, \aap, 605, A86, \dodoi{10.1051/0004-6361/201730807}

\bibitem[{{Manara} {et~al.}(2012){Manara}, {Robberto}, {Da Rio}, {Lodato}, {Hillenbrand}, {Stassun}, \& {Soderblom}}]{Manaraetal2012}
{Manara}, C.~F., {Robberto}, M., {Da Rio}, N., {et~al.} 2012, \apj, 755, 154, \dodoi{10.1088/0004-637X/755/2/154}

\bibitem[{{Manara} {et~al.}(2013{\natexlab{b}}){Manara}, {Testi}, {Rigliaco}, {Alcal{\'a}}, {Natta}, {Stelzer}, {Biazzo}, {Covino}, {Covino}, {Cupani}, {D'Elia}, \& {Randich}}]{Manaraetal2013}
{Manara}, C.~F., {Testi}, L., {Rigliaco}, E., {et~al.} 2013{\natexlab{b}}, \aap, 551, A107, \dodoi{10.1051/0004-6361/201220921}

\bibitem[{{Manara} {et~al.}(2021){Manara}, {Frasca}, {Venuti}, {Siwak}, {Herczeg}, {Calvet}, {Hernandez}, {Tychoniec}, {Gangi}, {Alcal{\'a}}, {Boffin}, {Nisini}, {Robberto}, {Briceno}, {Campbell-White}, {Sicilia-Aguilar}, {McGinnis}, {Fedele}, {K{\'o}sp{\'a}l}, {{\'A}brah{\'a}m}, {Alonso-Santiago}, {Antoniucci}, {Arulanantham}, {Bacciotti}, {Banzatti}, {Beccari}, {Benisty}, {Biazzo}, {Bouvier}, {Cabrit}, {Caratti o Garatti}, {Coffey}, {Covino}, {Dougados}, {Eisl{\"o}ffel}, {Ercolano}, {Espaillat}, {Erkal}, {Facchini}, {Fang}, {Fiorellino}, {Fischer}, {France}, {Gameiro}, {Garcia Lopez}, {Giannini}, {Ginski}, {Grankin}, {G{\"u}nther}, {Hartmann}, {Hillenbrand}, {Hussain}, {James}, {Koutoulaki}, {Lodato}, {Mauc{\'o}}, {Mendigut{\'\i}a}, {Mentel}, {Miotello}, {Oudmaijer}, {Rigliaco}, {Rosotti}, {Sanchis}, {Schneider}, {Spina}, {Stelzer}, {Testi}, {Thanathibodee}, {Vink}, {Walter}, {Williams}, \& {Zsidi}}]{manara2021}
{Manara}, C.~F., {Frasca}, A., {Venuti}, L., {et~al.} 2021, \aap, 650, A196, \dodoi{10.1051/0004-6361/202140639}

\bibitem[{{Mathieu} {et~al.}(1997){Mathieu}, {Stassun}, {Basri}, {Jensen}, {Johns-Krull}, {Valenti}, \& {Hartmann}}]{Mathieuetal1997}
{Mathieu}, R.~D., {Stassun}, K., {Basri}, G., {et~al.} 1997, \aj, 113, 1841, \dodoi{10.1086/118395}

\bibitem[{{McCully} {et~al.}(2018){McCully}, {Volgenau}, {Harbeck}, {Lister}, {Saunders}, {Turner}, {Siiverd}, \& {Bowman}}]{McCullyetal2018}
{McCully}, C., {Volgenau}, N.~H., {Harbeck}, D.-R., {et~al.} 2018, in Society of Photo-Optical Instrumentation Engineers (SPIE) Conference Series, Vol. 10707, \procspie, 107070K, \dodoi{10.1117/12.2314340}

\bibitem[{{Meijerink} {et~al.}(2009){Meijerink}, {Pontoppidan}, {Blake}, {Poelman}, \& {Dullemond}}]{meijerink09}
{Meijerink}, R., {Pontoppidan}, K.~M., {Blake}, G.~A., {Poelman}, D.~R., \& {Dullemond}, C.~P. 2009, \apj, 704, 1471, \dodoi{10.1088/0004-637X/704/2/1471}

\bibitem[{{Miotello} {et~al.}(2023){Miotello}, {Kamp}, {Birnstiel}, {Cleeves}, \& {Kataoka}}]{Miotello2023}
{Miotello}, A., {Kamp}, I., {Birnstiel}, T., {Cleeves}, L.~C., \& {Kataoka}, A. 2023, in Astronomical Society of the Pacific Conference Series, Vol. 534, Protostars and Planets VII, ed. S.~{Inutsuka}, Y.~{Aikawa}, T.~{Muto}, K.~{Tomida}, \& M.~{Tamura}, 501, \dodoi{10.48550/arXiv.2203.09818}

\bibitem[{{Modigliani} {et~al.}(2010){Modigliani}, {Goldoni}, {Royer}, {Haigron}, {Guglielmi}, {Fran{\c{c}}ois}, {Horrobin}, {Bristow}, {Vernet}, {Moehler}, {Kerber}, {Ballester}, {Mason}, \& {Christensen}}]{modigliani2010}
{Modigliani}, A., {Goldoni}, P., {Royer}, F., {et~al.} 2010, in Society of Photo-Optical Instrumentation Engineers (SPIE) Conference Series, Vol. 7737, Observatory Operations: Strategies, Processes, and Systems III, ed. D.~R. {Silva}, A.~B. {Peck}, \& B.~T. {Soifer}, 773728, \dodoi{10.1117/12.857211}

\bibitem[{{Mohanty} {et~al.}(2005){Mohanty}, {Jayawardhana}, \& {Basri}}]{Mohantyetal2005}
{Mohanty}, S., {Jayawardhana}, R., \& {Basri}, G. 2005, \apj, 626, 498, \dodoi{10.1086/429794}

\bibitem[{{Mu{\~n}oz} \& {Lai}(2016)}]{Munoz&Lai2016}
{Mu{\~n}oz}, D.~J., \& {Lai}, D. 2016, \apj, 827, 43, \dodoi{10.3847/0004-637X/827/1/43}

\bibitem[{{Mu{\~n}oz-Romero} {et~al.}(2024){Mu{\~n}oz-Romero}, {Banzatti}, {{\"O}berg}, \& et~al}]{romeromirza24b}
{Mu{\~n}oz-Romero}, C.~E., {Banzatti}, A., {{\"O}berg}, K.~I., \& et~al. 2024, \apj, submitted, \dodoi{...}

\bibitem[{{Muzerolle} {et~al.}(2019){Muzerolle}, {Flaherty}, {Balog}, {Beck}, \& {Gutermuth}}]{Muzerolleetal2019}
{Muzerolle}, J., {Flaherty}, K., {Balog}, Z., {Beck}, T., \& {Gutermuth}, R. 2019, \apj, 877, 29, \dodoi{10.3847/1538-4357/ab1756}

\bibitem[{{Natta} {et~al.}(2004){Natta}, {Testi}, {Muzerolle}, {Randich}, {Comer{\'o}n}, \& {Persi}}]{Nattaetal2004}
{Natta}, A., {Testi}, L., {Muzerolle}, J., {et~al.} 2004, \aap, 424, 603, \dodoi{10.1051/0004-6361:20040356}

\bibitem[{{Newville} {et~al.}(2014){Newville}, {Stensitzki}, {Allen}, \& {Ingargiola}}]{lmfit}
{Newville}, M., {Stensitzki}, T., {Allen}, D.~B., \& {Ingargiola}, A. 2014, {LMFIT: Non-Linear Least-Square Minimization and Curve-Fitting for Python}, 0.8.0,  Zenodo, \dodoi{10.5281/zenodo.11813}

\bibitem[{{{\"O}berg} \& {Bergin}(2021)}]{ObergBergin2021}
{{\"O}berg}, K.~I., \& {Bergin}, E.~A. 2021, \physrep, 893, 1, \dodoi{10.1016/j.physrep.2020.09.004}

\bibitem[{{{\"O}berg} {et~al.}(2023){{\"O}berg}, {Facchini}, \& {Anderson}}]{Oberg2023}
{{\"O}berg}, K.~I., {Facchini}, S., \& {Anderson}, D.~E. 2023, \araa, 61, 287, \dodoi{10.1146/annurev-astro-022823-040820}

\bibitem[{{Perotti} {et~al.}(2023){Perotti}, {Christiaens}, {Henning}, {Tabone}, {Waters}, {Kamp}, {Olofsson}, {Grant}, {Gasman}, {Bouwman}, {Samland}, {Franceschi}, {van Dishoeck}, {Schwarz}, {G{\"u}del}, {Lagage}, {Ray}, {Vandenbussche}, {Abergel}, {Absil}, {Arabhavi}, {Argyriou}, {Barrado}, {Boccaletti}, {Caratti o Garatti}, {Geers}, {Glauser}, {Justannont}, {Lahuis}, {Mueller}, {Nehm{\'e}}, {Pantin}, {Scheithauer}, {Waelkens}, {Guadarrama}, {Jang}, {Kanwar}, {Morales-Calder{\'o}n}, {Pawellek}, {Rodgers-Lee}, {Schreiber}, {Colina}, {Greve}, {{\"O}stlin}, \& {Wright}}]{perotti23}
{Perotti}, G., {Christiaens}, V., {Henning}, T., {et~al.} 2023, \nat, 620, 516, \dodoi{10.1038/s41586-023-06317-9}

\bibitem[{{Pontoppidan} {et~al.}(2024){Pontoppidan}, {Salyk}, {Banzatti}, {Zhang}, {Pascucci}, {{\"O}berg}, {Long}, {Mu{\~n}oz-Romero}, {Carr}, {Najita}, {Blake}, {Arulanantham}, {Andrews}, {Ballering}, {Bergin}, {Calahan}, {Cobb}, {Colmenares}, {Dickson-Vandervelde}, {Dignan}, {Green}, {Heretz}, {Herczeg}, {Kalyaan}, {Krijt}, {Pauly}, {Pinilla}, {Trapman}, \& {Xie}}]{pontoppidan24}
{Pontoppidan}, K.~M., {Salyk}, C., {Banzatti}, A., {et~al.} 2024, \apj, 963, 158, \dodoi{10.3847/1538-4357/ad20f0}

\bibitem[{{Price-Whelan} {et~al.}(2018){Price-Whelan}, {Sip{\H{o}}cz}, {G{\"u}nther}, {Lim}, {Crawford}, {Conseil}, {Shupe}, {Craig}, {Dencheva}, {Ginsburg}, {VanderPlas}, {Bradley}, {P{\'e}rez-Su{\'a}rez}, {de Val-Borro}, {Paper Contributors}, {Aldcroft}, {Cruz}, {Robitaille}, {Tollerud}, {Coordination Committee}, {Ardelean}, {Babej}, {Bach}, {Bachetti}, {Bakanov}, {Bamford}, {Barentsen}, {Barmby}, {Baumbach}, {Berry}, {Biscani}, {Boquien}, {Bostroem}, {Bouma}, {Brammer}, {Bray}, {Breytenbach}, {Buddelmeijer}, {Burke}, {Calderone}, {Cano Rodr{\'\i}guez}, {Cara}, {Cardoso}, {Cheedella}, {Copin}, {Corrales}, {Crichton}, {D{\textquoteright}Avella}, {Deil}, {Depagne}, {Dietrich}, {Donath}, {Droettboom}, {Earl}, {Erben}, {Fabbro}, {Ferreira}, {Finethy}, {Fox}, {Garrison}, {Gibbons}, {Goldstein}, {Gommers}, {Greco}, {Greenfield}, {Groener}, {Grollier}, {Hagen}, {Hirst}, {Homeier}, {Horton}, {Hosseinzadeh}, {Hu}, {Hunkeler}, {Ivezi{\'c}}, {Jain}, {Jenness}, {Kanarek}, {Kendrew}, {Kern}, {Kerzendorf}, {Khvalko},
  {King}, {Kirkby}, {Kulkarni}, {Kumar}, {Lee}, {Lenz}, {Littlefair}, {Ma}, {Macleod}, {Mastropietro}, {McCully}, {Montagnac}, {Morris}, {Mueller}, {Mumford}, {Muna}, {Murphy}, {Nelson}, {Nguyen}, {Ninan}, {N{\"o}the}, {Ogaz}, {Oh}, {Parejko}, {Parley}, {Pascual}, {Patil}, {Patil}, {Plunkett}, {Prochaska}, {Rastogi}, {Reddy Janga}, {Sabater}, {Sakurikar}, {Seifert}, {Sherbert}, {Sherwood-Taylor}, {Shih}, {Sick}, {Silbiger}, {Singanamalla}, {Singer}, {Sladen}, {Sooley}, {Sornarajah}, {Streicher}, {Teuben}, {Thomas}, {Tremblay}, {Turner}, {Terr{\'o}n}, {van Kerkwijk}, {de la Vega}, {Watkins}, {Weaver}, {Whitmore}, {Woillez}, {Zabalza}, \& {Contributors}}]{astropy1}
{Price-Whelan}, A.~M., {Sip{\H{o}}cz}, B.~M., {G{\"u}nther}, H.~M., {et~al.} 2018, \aj, 156, 123, \dodoi{10.3847/1538-3881/aabc4f}

\bibitem[{{Rieke} {et~al.}(2015){Rieke}, {Wright}, {B{\"o}ker}, {Bouwman}, {Colina}, {Glasse}, {Gordon}, {Greene}, {G{\"u}del}, {Henning}, {Justtanont}, {Lagage}, {Meixner}, {N{\o}rgaard-Nielsen}, {Ray}, {Ressler}, {van Dishoeck}, \& {Waelkens}}]{miri}
{Rieke}, G.~H., {Wright}, G.~S., {B{\"o}ker}, T., {et~al.} 2015, \pasp, 127, 584, \dodoi{10.1086/682252}

\bibitem[{{Rigliaco} {et~al.}(2015){Rigliaco}, {Pascucci}, {Duchene}, {Edwards}, {Ardila}, {Grady}, {Mendigut{\'\i}a}, {Montesinos}, {Mulders}, {Najita}, {Carpenter}, {Furlan}, {Gorti}, {Meijerink}, \& {Meyer}}]{rigliaco2015}
{Rigliaco}, E., {Pascucci}, I., {Duchene}, G., {et~al.} 2015, \apj, 801, 31, \dodoi{10.1088/0004-637X/801/1/31}

\bibitem[{{Robberto} {et~al.}(2004){Robberto}, {Song}, {Mora Carrillo}, {Beckwith}, {Makidon}, \& {Panagia}}]{Robbertoetal2004}
{Robberto}, M., {Song}, J., {Mora Carrillo}, G., {et~al.} 2004, \apj, 606, 952, \dodoi{10.1086/383141}

\bibitem[{{Salter} {et~al.}(2010){Salter}, {K{\'o}sp{\'a}l}, {Getman}, {Hogerheijde}, {van Kempen}, {Carpenter}, {Blake}, \& {Wilner}}]{Salteretal2010}
{Salter}, D.~M., {K{\'o}sp{\'a}l}, {\'A}., {Getman}, K.~V., {et~al.} 2010, \aap, 521, A32, \dodoi{10.1051/0004-6361/201015197}

\bibitem[{{Salyk} {et~al.}(2013){Salyk}, {Herczeg}, {Brown}, {Blake}, {Pontoppidan}, \& {van Dishoeck}}]{salyk13}
{Salyk}, C., {Herczeg}, G.~J., {Brown}, J.~M., {et~al.} 2013, \apj, 769, 21, \dodoi{10.1088/0004-637X/769/1/21}

\bibitem[{{Salyk} {et~al.}(2025){Salyk}, {Pontoppidan}, {Banzatti}, {Bergin}, {Arulanantham}, {Najita}, {Blake}, {Carr}, {Zhang}, \& {Xie}}]{salyk25}
{Salyk}, C., {Pontoppidan}, K.~M., {Banzatti}, A., {et~al.} 2025, arXiv e-prints, arXiv:2502.05061, \dodoi{10.48550/arXiv.2502.05061}

\bibitem[{{Schwarz} {et~al.}(2024){Schwarz}, {Henning}, {Christiaens}, {Gasman}, {Samland}, {Perotti}, {Jang}, {Grant}, {Tabone}, {Morales-Calder{\'o}n}, {Kamp}, {van Dishoeck}, {G{\"u}del}, {Lagage}, {Barrado}, {Caratti o Garatti}, {Glauser}, {Ray}, {Vandenbussche}, {Waters}, {Arabhavi}, {Kanwar}, {Olofsson}, {Rodgers-Lee}, {Schreiber}, \& {Temmink}}]{schwarz24}
{Schwarz}, K.~R., {Henning}, T., {Christiaens}, V., {et~al.} 2024, \apj, 962, 8, \dodoi{10.3847/1538-4357/ad1393}

\bibitem[{{Shu} {et~al.}(1994){Shu}, {Najita}, {Ostriker}, {Wilkin}, {Ruden}, \& {Lizano}}]{Shuetal1994}
{Shu}, F., {Najita}, J., {Ostriker}, E., {et~al.} 1994, \apj, 429, 781, \dodoi{10.1086/174363}

\bibitem[{{Sitko} {et~al.}(2023){Sitko}, {Russell}, {Long}, {Assani}, {Pikhartova}, {Bayyari}, {Grady}, {Lisse}, {Marengo}, {Wisniewski}, \& {Danchi}}]{Sitkoetal2023}
{Sitko}, M.~L., {Russell}, R.~W., {Long}, Z.~C., {et~al.} 2023, \aj, 166, 24, \dodoi{10.3847/1538-3881/acd7e8}

\bibitem[{{Smette} {et~al.}(2015){Smette}, {Sana}, {Noll}, {Horst}, {Kausch}, {Kimeswenger}, {Barden}, {Szyszka}, {Jones}, {Gallenne}, {Vinther}, {Ballester}, \& {Taylor}}]{molecfit}
{Smette}, A., {Sana}, H., {Noll}, S., {et~al.} 2015, \aap, 576, A77, \dodoi{10.1051/0004-6361/201423932}

\bibitem[{{Tabone} {et~al.}(2023){Tabone}, {Bettoni}, {van Dishoeck}, {Arabhavi}, {Grant}, {Gasman}, {Henning}, {Kamp}, {G{\"u}del}, {Lagage}, {Ray}, {Vandenbussche}, {Abergel}, {Absil}, {Argyriou}, {Barrado}, {Boccaletti}, {Bouwman}, {Caratti o Garatti}, {Geers}, {Glauser}, {Justannont}, {Lahuis}, {Mueller}, {Nehm{\'e}}, {Olofsson}, {Pantin}, {Scheithauer}, {Waelkens}, {Waters}, {Black}, {Christiaens}, {Guadarrama}, {Morales-Calder{\'o}n}, {Jang}, {Kanwar}, {Pawellek}, {Perotti}, {Perrin}, {Rodgers-Lee}, {Samland}, {Schreiber}, {Schwarz}, {Colina}, {{\"O}stlin}, \& {Wright}}]{Tabone2023}
{Tabone}, B., {Bettoni}, G., {van Dishoeck}, E.~F., {et~al.} 2023, Nature Astronomy, 7, 805, \dodoi{10.1038/s41550-023-01965-3}

\bibitem[{{Temmink} {et~al.}(2024){Temmink}, {van Dishoeck}, {Grant}, {Tabone}, {Gasman}, {Christiaens}, {Samland}, {Argyriou}, {Perotti}, {G{\"u}del}, {Henning}, {Lagage}, {Abergel}, {Absil}, {Barrado}, {Caratti o Garatti}, {Glauser}, {Kamp}, {Lahuis}, {Olofsson}, {Ray}, {Scheithauer}, {Vandenbussche}, {Waters}, {Arabhavi}, {Jang}, {Kanwar}, {Morales-Calder{\'o}n}, {Rodgers-Lee}, {Schreiber}, {Schwarz}, \& {Colina}}]{temmink24}
{Temmink}, M., {van Dishoeck}, E.~F., {Grant}, S.~L., {et~al.} 2024, \aap, 686, A117, \dodoi{10.1051/0004-6361/202348911}

\bibitem[{{Tofflemire} {et~al.}(2017){Tofflemire}, {Mathieu}, {Ardila}, {Akeson}, {Ciardi}, {Johns-Krull}, {Herczeg}, \& {Quijano-Vodniza}}]{Tofflemireetal2017a}
{Tofflemire}, B.~M., {Mathieu}, R.~D., {Ardila}, D.~R., {et~al.} 2017, \apj, 835, 8, \dodoi{10.3847/1538-4357/835/1/8}

\bibitem[{{Tychoniec} {et~al.}(2024){Tychoniec}, {van Gelder}, {van Dishoeck}, {Francis}, {Rocha}, {Caratti o Garatti}, {Beuther}, {Gieser}, {Justtanont}, {Linnartz}, {Le Gouellec}, {Perotti}, {Devaraj}, {Tabone}, {Ray}, {Brunken}, {Chen}, {Kavanagh}, {Klaassen}, {Slavicinska}, {G{\"u}del}, \& {{\"O}stlin}}]{Tychoniecetal2024}
{Tychoniec}, {\L}., {van Gelder}, M.~L., {van Dishoeck}, E.~F., {et~al.} 2024, \aap, 687, A36, \dodoi{10.1051/0004-6361/202348889}

\bibitem[{{Valenti} {et~al.}(1993){Valenti}, {Basri}, \& {Johns}}]{Valentietal1993}
{Valenti}, J.~A., {Basri}, G., \& {Johns}, C.~M. 1993, \aj, 106, 2024, \dodoi{10.1086/116783}

\bibitem[{{van Dishoeck} {et~al.}(2023){van Dishoeck}, {Grant}, {Tabone}, {van Gelder}, {Francis}, {Tychoniec}, {Bettoni}, {Arabhavi}, {Gasman}, {Nazari}, {Vlasblom}, {Kavanagh}, {Christiaens}, {Klaassen}, {Beuther}, {Henning}, \& {Kamp}}]{vanDishoeck2023}
{van Dishoeck}, E.~F., {Grant}, S., {Tabone}, B., {et~al.} 2023, Faraday Discussions, 245, 52, \dodoi{10.1039/D3FD00010A}

\bibitem[{{Venuti} {et~al.}(2014){Venuti}, {Bouvier}, {Flaccomio}, {Alencar}, {Irwin}, {Stauffer}, {Cody}, {Teixeira}, {Sousa}, {Micela}, {Cuillandre}, \& {Peres}}]{Venutietal2014}
{Venuti}, L., {Bouvier}, J., {Flaccomio}, E., {et~al.} 2014, \aap, 570, A82, \dodoi{10.1051/0004-6361/201423776}

\bibitem[{{Vernet} {et~al.}(2011){Vernet}, {Dekker}, {D'Odorico}, {Kaper}, {Kjaergaard}, {Hammer}, {Randich}, {Zerbi}, {Groot}, {Hjorth}, {Guinouard}, {Navarro}, {Adolfse}, {Albers}, {Amans}, {Andersen}, {Andersen}, {Binetruy}, {Bristow}, {Castillo}, {Chemla}, {Christensen}, {Conconi}, {Conzelmann}, {Dam}, {de Caprio}, {de Ugarte Postigo}, {Delabre}, {di Marcantonio}, {Downing}, {Elswijk}, {Finger}, {Fischer}, {Flores}, {Fran{\c{c}}ois}, {Goldoni}, {Guglielmi}, {Haigron}, {Hanenburg}, {Hendriks}, {Horrobin}, {Horville}, {Jessen}, {Kerber}, {Kern}, {Kiekebusch}, {Kleszcz}, {Klougart}, {Kragt}, {Larsen}, {Lizon}, {Lucuix}, {Mainieri}, {Manuputy}, {Martayan}, {Mason}, {Mazzoleni}, {Michaelsen}, {Modigliani}, {Moehler}, {M{\o}ller}, {Norup S{\o}rensen}, {N{\o}rregaard}, {P{\'e}roux}, {Patat}, {Pena}, {Pragt}, {Reinero}, {Rigal}, {Riva}, {Roelfsema}, {Royer}, {Sacco}, {Santin}, {Schoenmaker}, {Spano}, {Sweers}, {Ter Horst}, {Tintori}, {Tromp}, {van Dael}, {van der Vliet}, {Venema}, {Vidali}, {Vinther}, {Vola},
  {Winters}, {Wistisen}, {Wulterkens}, \& {Zacchei}}]{vernet11}
{Vernet}, J., {Dekker}, H., {D'Odorico}, S., {et~al.} 2011, \aap, 536, A105, \dodoi{10.1051/0004-6361/201117752}

\bibitem[{{Virtanen} {et~al.}(2020){Virtanen}, {Gommers}, {Oliphant}, {Haberland}, {Reddy}, {Cournapeau}, {Burovski}, {Peterson}, {Weckesser}, {Bright}, {van der Walt}, {Brett}, {Wilson}, {Jarrod Millman}, {Mayorov}, {Nelson}, {Jones}, {Kern}, {Larson}, {Carey}, {Polat}, {Feng}, {Moore}, {Vand erPlas}, {Laxalde}, {Perktold}, {Cimrman}, {Henriksen}, {Quintero}, {Harris}, {Archibald}, {Ribeiro}, {Pedregosa}, {van Mulbregt}, \& {Contributors}}]{scipy2}
{Virtanen}, P., {Gommers}, R., {Oliphant}, T.~E., {et~al.} 2020, Nature Methods, 17, 261, \dodoi{https://doi.org/10.1038/s41592-019-0686-2}

\bibitem[{{Vlasblom} {et~al.}(2024){Vlasblom}, {Temmink}, {Grant}, {Kurtovic}, {Sellek}, {van Dishoeck}, {G{\"u}del}, {Henning}, {Lagage}, {Barrado}, {Garatti}, {Glauser}, {Kamp}, {Lahuis}, {Olofsson}, {Arabhavi}, {Christiaens}, {Gasman}, {Jang}, {Morales-Calder{\'o}n}, {Perotti}, {Schwarz}, \& {Tabone}}]{vlasblom24}
{Vlasblom}, M., {Temmink}, M., {Grant}, S.~L., {et~al.} 2024, arXiv e-prints, arXiv:2412.12715, \dodoi{10.48550/arXiv.2412.12715}

\bibitem[{{Wells} {et~al.}(2015){Wells}, {Pel}, {Glasse}, {Wright}, {Aitink-Kroes}, {Azzollini}, {Beard}, {Brandl}, {Gallie}, {Geers}, {Glauser}, {Hastings}, {Henning}, {Jager}, {Justtanont}, {Kruizinga}, {Lahuis}, {Lee}, {Martinez-Delgado}, {Mart{\'\i}nez-Galarza}, {Meijers}, {Morrison}, {M{\"u}ller}, {Nakos}, {O'Sullivan}, {Oudenhuysen}, {Parr-Burman}, {Pauwels}, {Rohloff}, {Schmalzl}, {Sykes}, {Thelen}, {van Dishoeck}, {Vandenbussche}, {Venema}, {Visser}, {Waters}, \& {Wright}}]{jwst-mrs}
{Wells}, M., {Pel}, J.~W., {Glasse}, A., {et~al.} 2015, \pasp, 127, 646, \dodoi{10.1086/682281}

\bibitem[{{Woitke} {et~al.}(2024){Woitke}, {Thi}, {Arabhavi}, {Kamp}, {K{\'o}sp{\'a}l}, \& {{\'A}brah{\'a}m}}]{Woitke2024}
{Woitke}, P., {Thi}, W.~F., {Arabhavi}, A.~M., {et~al.} 2024, \aap, 683, A219, \dodoi{10.1051/0004-6361/202347730}

\bibitem[{{Wright} {et~al.}(2023){Wright}, {Rieke}, {Glasse}, {Ressler}, {Garc{\'\i}a Mar{\'\i}n}, {Aguilar}, {Alberts}, {{\'A}lvarez-M{\'a}rquez}, {Argyriou}, {Banks}, {Baudoz}, {Boccaletti}, {Bouchet}, {Bouwman}, {Brandl}, {Breda}, {Bright}, {Cale}, {Colina}, {Cossou}, {Coulais}, {Cracraft}, {De Meester}, {Dicken}, {Engesser}, {Etxaluze}, {Fox}, {Friedman}, {Fu}, {Gasman}, {G{\'a}sp{\'a}r}, {Gastaud}, {Geers}, {Glauser}, {Gordon}, {Greene}, {Greve}, {Grundy}, {G{\"u}del}, {Guillard}, {Haderlein}, {Hashimoto}, {Henning}, {Hines}, {Holler}, {Detre}, {Jahromi}, {James}, {Jones}, {Justtanont}, {Kavanagh}, {Kendrew}, {Klaassen}, {Krause}, {Labiano}, {Lagage}, {Lambros}, {Larson}, {Law}, {Lee}, {Libralato}, {Lorenzo Alverez}, {Meixner}, {Morrison}, {Mueller}, {Murray}, {Mycroft}, {Myers}, {Nayak}, {Naylor}, {Nickson}, {Noriega-Crespo}, {{\"O}stlin}, {O'Sullivan}, {Ottens}, {Patapis}, {Penanen}, {Pietraszkiewicz}, {Ray}, {Regan}, {Roteliuk}, {Royer}, {Samara-Ratna}, {Samuelson}, {Sargent}, {Scheithauer},
  {Schneider}, {Schreiber}, {Shaughnessy}, {Sheehan}, {Shivaei}, {Sloan}, {Tamas}, {Teague}, {Temim}, {Tikkanen}, {Tustain}, {van Dishoeck}, {Vandenbussche}, {Weilert}, {Whitehouse}, \& {Wolff}}]{miri2}
{Wright}, G.~S., {Rieke}, G.~H., {Glasse}, A., {et~al.} 2023, \pasp, 135, 048003, \dodoi{10.1088/1538-3873/acbe66}

\bibitem[{{Zsidi} {et~al.}(2022){Zsidi}, {Manara}, {K{\'o}sp{\'a}l}, {Hussain}, {{\'A}brah{\'a}m}, {Alecian}, {B{\'o}di}, {P{\'a}l}, \& {Sarkis}}]{Zsidietal2022}
{Zsidi}, G., {Manara}, C.~F., {K{\'o}sp{\'a}l}, {\'A}., {et~al.} 2022, \aap, 660, A108, \dodoi{10.1051/0004-6361/202142203}

\end{thebibliography}

\end{document}